\documentclass[floatfix,showpacs,amsmath,amssymb,twocolumn,aps,prapplied,10pt,longbibliography,superscriptaddress,notitlepage]{revtex4-2}
\usepackage[T1]{fontenc}
\usepackage{comment}
\usepackage{textcomp}
\usepackage{amsmath,amssymb,amsthm,mathrsfs,dsfont}
\usepackage{graphicx}
\usepackage[svgnames,psnames]{xcolor}
\usepackage{hyperref}
\hypersetup{colorlinks,citecolor=DarkGreen,linkcolor=FireBrick,urlcolor=FireBrick,linktocpage,unicode}
\usepackage{braket}
\usepackage{bm}
\usepackage{here}
\usepackage{cancel}

\allowdisplaybreaks[2]
\newcommand{\TAred}[1]{\textcolor{black}{#1}}
\begin{document}
\title{Control of the \texorpdfstring{$ZZ$}{ZZ} coupling between Kerr-cat qubits via transmon couplers}

\author{Takaaki Aoki}
\email{takaaki-aoki@aist.go.jp}
\affiliation{Global Research and Development Center for Business by Quantum-AI Technology (G-QuAT), National Institute of Advanced Industrial Science and Technology (AIST), 1-1-1 Umezono, Tsukuba, Ibaraki 305-8568, Japan}

\author{Taro Kanao}
\affiliation{Corporate Research and Development Center, Toshiba Corporation, 1, Komukai-Toshiba-cho, Saiwai-ku, Kawasaki, Kanagawa 212-8582, Japan}

\author{Hayato Goto}
\affiliation{Corporate Research and Development Center, Toshiba Corporation, 1, Komukai-Toshiba-cho, Saiwai-ku, Kawasaki, Kanagawa 212-8582, Japan}

\author{Shiro Kawabata}
\affiliation{Global Research and Development Center for Business by Quantum-AI Technology (G-QuAT), National Institute of Advanced Industrial Science and Technology (AIST), 1-1-1 Umezono, Tsukuba, Ibaraki 305-8568, Japan}
\affiliation{NEC-AIST Quantum Technology Cooperative Research Laboratory, National Institute of Advanced Industrial Science and Technology (AIST), 1-1-1 Umezono, Tsukuba, Ibaraki 305-8568, Japan}

\author{Shumpei Masuda}
\email{shumpei.masuda@aist.go.jp}
\affiliation{Global Research and Development Center for Business by Quantum-AI Technology (G-QuAT), National Institute of Advanced Industrial Science and Technology (AIST), 1-1-1 Umezono, Tsukuba, Ibaraki 305-8568, Japan}
\affiliation{NEC-AIST Quantum Technology Cooperative Research Laboratory, National Institute of Advanced Industrial Science and Technology (AIST), 1-1-1 Umezono, Tsukuba, Ibaraki 305-8568, Japan}

\begin{abstract}
Kerr-cat qubits are a promising candidate for fault-tolerant quantum computers owing to the biased nature of their errors.
The $ZZ$ coupling between the qubits can be utilized for a two-qubit entangling gate, but the residual coupling \TAred{called $ZZ$ crosstalk is detrimental to precise computing}.
In order to resolve this problem, we propose a tunable $ZZ$-coupling scheme using two transmon couplers.
By setting the detunings of the two  couplers at opposite values, the residual $ZZ$ couplings via the two couplers cancel each other out.
We also apply our scheme to the $R_{zz}(\Theta)$ gate ($ZZ$ rotation with angle $\Theta$), one of the two-qubit entangling gates.
We numerically show that the fidelity of the $R_{zz}(-\pi/2)$ gate is higher than 99.9\% in a case of $16$-ns gate time and without decoherence.
\end{abstract}
\maketitle
\section{Introduction}
Quantum computation \cite{nielsen_chuang_2010} is expected to surpass classical computation in speed in specific problems such as prime factorization \cite{doi:10.1137/S0036144598347011}, database search \cite{10.1145/237814.237866}, quantum chemistry \cite{RevModPhys.92.015003}, and machine learning \cite{Biamonte2017}.
A major obstacle is noise caused by the interaction between a computing system and its environment \cite{schlosshauer2007decoherence,SCHLOSSHAUER20191}.
In order to cancel out the noise and obtain reliable computational results, we must perform quantum error correction \cite{Devitt_2013,doi:10.1080/00107514.2019.1667078}.
This requires a considerable overhead cost, which makes difficult the construction of a large-scale fault-tolerant quantum computer \cite[Sec.~3.2]{Preskill2018quantumcomputingin}.

However, when the noise is biased, we can reduce the overhead \cite{PhysRevA.92.062309}.
As such \cite{doi:10.1126/sciadv.aay5901_2020_8_21,PhysRevApplied.18.024076,PhysRevResearch.4.013082_2022_2_2}, a Kerr-cat qubit, which uses two coherent states with opposite phases as logical states and whose bit-flip error is exponentially suppressed with its photon number \cite[Sec.~I]{PhysRevX.9.041009_2019_10_9} \cite[Sec.~S2]{Suzuki2023}, has attracted much attention in recent years \cite{PhysRevX.9.041009_2019_10_9,PhysRevApplied.18.024076,PhysRevA.59.2631,PhysRevA.93.050301,Puri2017,Grimm2020_8_1,PhysRevApplied.10.024019_2018_8_15,Kanao2021_1_29,TomohiroYAMAJI20222021SEP0006,doi:10.1126/sciadv.1602273,Onodera2020_5_29,Goto2020_12_18,Goto2016,Puri2017_Jun_08,doi:10.7566/JPSJ.88.061015_2019_3_1,Kewming_2020_5_26,doi:10.1126/sciadv.aay5901_2020_8_21,Goto2018_5_8,PRXQuantum.2.030345_2021_9_16,PhysRevApplied.18.034076,PhysRevResearch.4.043054,2023arXiv231020108K,Masuda2021,PhysRevResearch.4.013082_2022_2_2,PhysRevLett.128.110502_2022_3_14,PhysRevResearch.4.013233_2022_3,PhysRevApplied.18.014019_2022_7,Suzuki2023}.
Universal gate sets \cite{PhysRevA.59.2631, PhysRevA.93.050301,Puri2017} and bias-preserving gates \cite{doi:10.1126/sciadv.aay5901_2020_8_21,PhysRevApplied.18.024076,PhysRevResearch.4.013082_2022_2_2} for Kerr-cat qubits have been theoretically proposed.
A full set of single-qubit gates on a Kerr-cat qubit has been performed experimentally \cite{Grimm2020_8_1}.
Quantum annealing \cite{PhysRevApplied.10.024019_2018_8_15,Goto2016,doi:10.1126/sciadv.1602273,Puri2017_Jun_08,doi:10.7566/JPSJ.88.061015_2019_3_1,Kewming_2020_5_26,Onodera2020_5_29,Goto2020_12_18,Kanao2021_1_29,TomohiroYAMAJI20222021SEP0006} and Boltzmann sampling \cite{Goto2018_5_8} based on Kerr-cat qubits have also been studied theoretically.

A Kerr-cat qubit is realized by a Kerr parametric oscillator (KPO), which is an oscillator under parametric (squeezing) drive with Kerr nonlinearity larger than the single-photon loss rate (single-photon Kerr regime) \cite{PhysRevA.59.2631,Goto2016,Puri2017,doi:10.7566/JPSJ.88.061015_2019_3_1,PhysRevA.105.023519}.
A typical coupling between KPOs is the beam-splitter type \cite{doi:10.1126/sciadv.aay5901_2020_8_21,PhysRevApplied.18.024076,Goto2016,PhysRevA.93.050301,doi:10.1126/sciadv.1602273,Puri2017,Puri2017_Jun_08,Goto2018_5_8, doi:10.7566/JPSJ.88.061015_2019_3_1,Kewming_2020_5_26,Onodera2020_5_29,Goto2020_12_18,PRXQuantum.2.030345_2021_9_16,PhysRevApplied.18.034076,PhysRevResearch.4.043054,2023arXiv231020108K,PhysRevApplied.20.014057}.
This leads to the $ZZ$ (longitudinal) coupling between Kerr-cat qubits \cite{doi:10.1126/sciadv.aay5901_2020_8_21,Goto2016,PhysRevA.93.050301,Puri2017,Puri2017_Jun_08,Goto2018_5_8,doi:10.7566/JPSJ.88.061015_2019_3_1,Kewming_2020_5_26,PRXQuantum.2.030345_2021_9_16,PhysRevApplied.18.034076,PhysRevResearch.4.043054,2023arXiv231020108K}, which can be utilized for an $R_{zz}$ ($ZZ$-rotation) gate \cite{doi:10.1126/sciadv.aay5901_2020_8_21,PhysRevA.93.050301,Puri2017,doi:10.7566/JPSJ.88.061015_2019_3_1,PhysRevApplied.18.034076,PhysRevResearch.4.043054,2023arXiv231020108K}, one of the two-qubit entangling gates.
The serious practical problem is that 
\TAred{even when we do not intend to perform an $R_{zz}$ gate, there are generally residual $ZZ$ couplings, called $ZZ$ crosstalk \cite{9193969,PhysRevApplied.12.054023_2019_11,PhysRevApplied.14.024070_2020_8,PhysRevLett.125.200503,PhysRevLett.125.200504_2020_11,PhysRevLett.127.080505_2021_8,PhysRevApplied.16.064062_2021_12,PhysRevApplied.18.034038_2022_9,PRXQuantum.4.010314,PRXQuantum.3.020301,PhysRevApplied.18.024068,PhysRevLett.129.060501}.
The crosstalk creates unwanted correlations between qubits \cite{Sarovar2020detectingcrosstalk}; an operation on one qubit may give unexpected effects on another one (violation of locality); an operation may be affected by another simultaneous one (violation of independence).
Since crosstalk prevents precise computing in this way, it has been studied extensively: its characterization \cite{9193969,PhysRevLett.126.230502,PRXQuantum.2.040338}; its detection \cite{Sarovar2020detectingcrosstalk}; its impact on simultaneous gate operations \cite{PRXQuantum.3.020301}; its suppression using tunable couplers \cite{PhysRevLett.113.220502_2014_11,PhysRevApplied.10.054062_2018_11,PhysRevApplied.12.054023_2019_11,PhysRevApplied.14.024070_2020_8,PhysRevLett.125.200503,PhysRevLett.125.200504_2020_11,PhysRevLett.127.080505_2021_8,PhysRevApplied.16.064062_2021_12,PhysRevApplied.18.034038_2022_9,PRXQuantum.4.010314}, dynamical decoupling \cite{PhysRevApplied.18.024068}, and multicolor drives \cite{PhysRevLett.129.060501}.}

In other systems such as transmons, tunable couplers have been used to eliminate unwanted residual couplings \cite{PhysRevLett.113.220502_2014_11,PhysRevApplied.10.054062_2018_11,PhysRevApplied.12.054023_2019_11,PhysRevApplied.14.024070_2020_8,PhysRevLett.125.200503,PhysRevLett.125.200504_2020_11,PhysRevLett.127.080505_2021_8,PhysRevApplied.16.064062_2021_12,PhysRevApplied.18.034038_2022_9,PRXQuantum.4.010314}, but not yet in KPO systems. 
In this paper, we propose a tunable $ZZ$-coupling scheme between two Kerr-cat qubits using two transmon couplers.
In this scheme, the detuning of one of the couplers is modulated to control the amplitude of the effective $ZZ$ coupling between the qubits.
Of note, the residual coupling can be eliminated when the detunings of the two couplers are set to opposite values.

Our scheme does not utilize resonance-frequency difference between KPOs unlike the cross-resonance gate \cite{PhysRevB.74.140504_2006_10,PhysRevB.81.134507_2010_4,PhysRevA.93.060302_2016_6,PhysRevA.102.042605_2020_10}, and thus can be implemented with identical KPOs. Therefore, it is expected that our scheme will reduce the complexity of design and fabrication of multi-KPO systems.
Our scheme can also mitigate the frequency-crowding problem, which becomes significant especially when frequency differences between qubits are needed \cite{9251858}. In addition, our scheme is compatible with KPO systems with lattice structures in contrast to the previous one which utilizes the phase difference between pump fields and can be applied only to simple KPO networks such as one-dimensional chains \cite{PhysRevApplied.18.034076,PhysRevApplied.20.014057}. Therefore, our scheme is advantageous for system scale up.

This article is organized as follows.
In Sec.~\ref{sec:Settings}, we introduce our system Hamiltonian and logical states of Kerr-cat qubits.
In Sec.~\ref{sec:coupling}, we explain how to switch {\scshape on} and {\scshape off} the $ZZ$ coupling between the qubits.
In Sec.~\ref{sec:gate}, we numerically evaluate the residual coupling and the $R_{zz}(-\pi/2)$-gate fidelity.
Finally, we conclude this article in Sec.~\ref{sec:conclusion}.

\section{Settings}
\label{sec:Settings}
\begin{figure}
    \centering
    \includegraphics[width=0.25\textwidth]{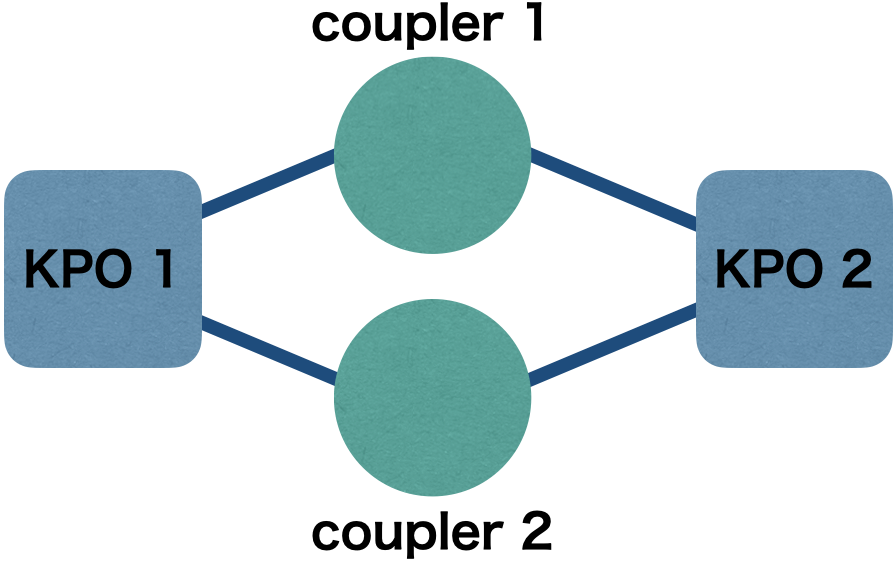}
    \caption{\TAred{A} schematic diagram of the studied system.
    Two identical KPOs are coupled via two transmon couplers.
    Our arrangement is similar to that in Ref.~\cite[Fig.~1(a)]{PhysRevApplied.12.054023_2019_11}.
    \TAred{An example of a circuit to realize the system is given in Appendix \ref{app:Circuit model}.}}
    \label{fig:system1}
\end{figure}
We consider a system consisting of two identical KPOs \TAred{(KPO $1$ and KPO $2$)} and two transmon couplers \TAred{(coupler $1$ and coupler $2$)}; see Fig.~\ref{fig:system1}.
The KPOs are parametrically driven with frequency $\omega_p$, which is twice their dressed resonance frequency.
The effective Hamiltonian of the system \TAred{$\hat{H}^{\mathrm{R}}(t)$} in a
rotating frame at frequency $\omega_p/2$ is written, in rotating-wave approximation, as
\begin{align}
    \hat{H}^{\mathrm{R}}(t)&=\sum_{j=1}^2\hat{H}_{\mathrm{KPO}j}^{\mathrm{R}}
    +\sum_{k=1}^2\hat{H}_{\mathrm{c}k}^{\mathrm{R}}(t)+\hat{H}_I^{\mathrm{R}},
    \label{eq:H_R_t_main1} \\
    \hat{H}_{\mathrm{KPO}j}^{\mathrm{R}}\TAred{/\hbar}&=-\frac{K}{2}\hat{a}_j^{\dag2}\hat{a}_j^2+\frac{p}{2}\left(\hat{a}_j^{\dag2}+\hat{a}_j^2\right),
    \label{eq:H_KPOj_R_main1} \\
    \hat{H}_{\mathrm{c}k}^{\mathrm{R}}(t)\TAred{/\hbar}&=
    -\frac{\chi_k(t)}{2}\hat{c}_k^{\dagger2}\hat{c}_k^2+\Delta_k(t)\hat{c}_k^{\dagger}\hat{c}_k,
    \label{eq:H_ck_R_t_main1} \\
    \hat{H}_I^{\mathrm{R}}\TAred{/\hbar}&=\sum_{j,k=1}^2g_{j,k}\left(\hat{a}_j\hat{c}_k^{\dagger}+\hat{a}_j^{\dagger}\hat{c}_k\right),
    \label{eq:H_I_R_main1}
\end{align}
where $\hat{H}_{\mathrm{KPO}j}^{\mathrm{R}}$, $\hat{H}_{\mathrm{c}k}^{\mathrm{R}}(t)$, and $\hat{H}_I^{\mathrm{R}}$ are the Hamiltonian of KPO $j$, the Hamiltonian of coupler $k$, and the interaction Hamiltonian, respectively;
\TAred{$\hbar=h/(2\pi)$ is the reduced Planck constant;}
$\hat{a}_j$ and $\hat{c}_k$ are the annihilation operators of KPO $j$ and coupler $k$;
$K(>0)$ and $p(>0)$ are the Kerr nonlinearity parameter and amplitude of the parametric drive of the KPOs;
$\chi_{k}\TAred{(t)}$ and $\Delta_{k}(t)$ are the Kerr nonlinearity parameter and detuning of coupler $k$;
$\Delta_1(t):=\omega_{\mathrm{c1}}(t)-\omega_p/2$ is time dependent but $\Delta_2:=\omega_{\mathrm{c2}}-\omega_p/2$ is not, where $\omega_{\mathrm{c1}}(t)$ ($\omega_{\mathrm{c2}}$) is a tunable (fixed) resonance frequency of coupler $1$ ($2$);
$g_{j,k}$ is the coupling strength between KPO $j$ and coupler $k$.
\TAred{A circuit to realize the above system is given in Appendix \ref{app:Circuit model}.}
We assume \TAred{positive constant coefficients such} that $\chi_{k}\TAred{(t)}=\chi >0$ and $g_{j,k}=g>0$ throughout \TAred{the main text} for simplicity.

We can transform $\hat{H}^{\mathrm{R}}_{\mathrm{KPO}j}$ as
\begin{align}
    \hat{H}^{\mathrm{R}}_{\mathrm{KPO}j}\TAred{/\hbar}=
    -\frac{K}{2}
    \left(\hat{a}_j^{\dagger2}
    -\alpha^2\right)
    \left(\hat{a}_j^2
    -\alpha^2\right)
    +\frac{K\alpha^4}{2}
\end{align}
with $\alpha=\sqrt{p/K}$, which shows that two coherent states $\ket{\pm\alpha}_{\mathrm{KPO}j}$ are the doubly degenerate highest levels of KPO $j$.
We assume that $\alpha$ is  sufficiently large so that the overlap between the coherent states ${\vphantom{\braket{\alpha|-\alpha}}}_{\mathrm{KPO}j}\braket{\alpha|-\alpha}_{\mathrm{KPO}j}=\mathrm{e}^{-2\alpha^2}$ is negligible.
It is known that these coherent states are stable
in the sense that their life time is an exponential function of $\alpha^2$ \cite[Sec.~I]{PhysRevX.9.041009_2019_10_9} \cite[Sec.~S2]{Suzuki2023}.
We use these coherent states to encode logical Kerr-cat qubits:
\begin{align}
    \ket{\bar{0}}_{\mathrm{q}j}:=\ket{\alpha}_{\mathrm{KPO}j},\quad
    \ket{\bar{1}}_{\mathrm{q}j}:=\ket{-\alpha}_{\mathrm{KPO}j}.
    \label{eq:basis1}
\end{align}
Then,
$\ket{\bar{0},\bar{0}}_{\rm q}:=\ket{\bar{0}}_{\mathrm{q}1}\otimes\ket{\bar{0}}_{\mathrm{q}2}$, $\ket{\bar{0},\bar{1}}_{\rm q}$, $\ket{\bar{1},\bar{0}}_{\rm q}$, and $\ket{\bar{1},\bar{1}}_{\rm q}$ form a complete set of basis states of the two Kerr-cat qubits.

\section{Switching of \texorpdfstring{$ZZ$}{ZZ} coupling}
\label{sec:coupling}
Let us explain briefly the mechanism behind the effective $ZZ$ coupling between the Kerr-cat qubits, which is controlled through $\Delta_1(t)$.
When $\Delta_1(t)=-\Delta_2$ and $|\Delta_2|\gg g\alpha$, as explained later, four states, $\ket{\bar{i},\bar{j}}_{\mathrm{q}}\otimes\ket{0,0}_{\mathrm{c}}$ ($i,j\in\{0,1\}$), are almost degenerate and the effective coupling between the qubits \TAred{is} suppressed.
When $\Delta_1(t)\neq-\Delta_2$, the effective coupling is on.
We gradually modulate $\Delta_1(t)$ in order to perform an $R_{zz}$ gate.
During the modulation, the states of the Kerr-cat qubits remain $\ket{\bar{i},\bar{j}}_{\rm q}$, as verified in \TAred{Appendix \ref{app:qubits_states}};
the degeneracy between the two sets of levels corresponding to $\{\ket{\bar{0},\bar{0}}_\mathrm{q},\ket{\bar{1},\bar{1}}_\mathrm{q}\}$ and $\{\ket{\bar{0},\bar{1}}_\mathrm{q},\ket{\bar{1},\bar{0}}_\mathrm{q}\}$ is lifted.
The energy difference between the two sets of levels can be regarded as the effect of the $ZZ$-coupling Hamiltonian of the form  $E\hat{\sigma}_{z}^{\mathrm{q}1}\hat{\sigma}_{z}^{\mathrm{q}2}\TAred{/2}$, where
\TAred{$\hat{\sigma}_{z}^{\mathrm{q}j}=\ket{\bar{0}}_{\mathrm{q}j}\bra{\bar{0}}-\ket{\bar{1}}_{\mathrm{q}j}\bra{\bar{1}}$} and $E$ is the energy difference.

In order to understand the above mechanism in detail, it is useful to consider the effective Hamiltonians of the couplers conditioned by the state of the Kerr-cat qubits which is assumed to be either of $\ket{\bar{i},\bar{j}}_{\rm q}$ ($i,j\in\{0,1\}$).
The effective Hamiltonian corresponding to $\ket{\bar{i},\bar{j}}_{\rm q}$ is defined as $\TAred{\hat{H}}_{\mathrm{c}}^{\TAred{\mathrm{R},}\Bar{i},\Bar{j}}(t)={}_{\rm q}\langle{\Bar{i},\Bar{j}}|{\hat{H}\TAred{^{\mathrm{R}}}(t)}|
    {\Bar{i},\Bar{j}}\rangle_{\rm q}$.
Using Eq.~\eqref{eq:basis1}, we obtain 
\begin{align}
\hat{H}_{\mathrm{c}}^{\TAred{\mathrm{R},}\Bar{i},\Bar{i}}(t)\TAred{/\hbar}
&=\sum_{k=1}^2
\left\{-\frac{\chi}{2}
\hat{c}_{k}^{\dagger2}
\hat{c}_{k}^2
+\Delta_{k}(t)
\left[\hat{c}_{k}^{\dagger}
+(-1)^{i}\alpha_{k}(t)\right]
\right.
\notag \\
&\quad\left.\times\left[\hat{c}_{k}+(-1)^{i}\alpha_{k}(t)\right]\vphantom{\frac{\chi}{2}}-2g\alpha\alpha_{k}(t)\right\}+K\alpha^4,
\label{eq:H_eff_c_same_sign_1}
\end{align}
\begin{align}
\TAred{\hat{H}}_{\mathrm{c}}^{\TAred{\mathrm{R},}\Bar{i},\Bar{j}(\ne \Bar{i})}(t)\TAred{/\hbar}
&=\sum_{k=1}^2 \Big{[}-\frac{\chi}{2} \hat{c}_{k}^{\dagger2}
\hat{c}_{k}^2
+\Delta_{k}(t)\hat{c}_{k}^{\dagger}\hat{c}_{k} \Big{]}
+K\alpha^4,
\label{eq:H_eff_c_different_sign_1}
\end{align}
where $\alpha_{k}(t)=2g\alpha/\Delta_{k}(t)$.
Note that $\alpha_2=2g\alpha/\Delta_{2}$ is time independent.
When the nonlinear terms in Eq.~\eqref{eq:H_eff_c_same_sign_1} can be neglected, the tensor product of coherent states $\ket{(-1)^{i+1}\alpha_{1}(t),(-1)^{i+1}\alpha_{2}}_{\rm c}$ is the eigenstate of Hamiltonian $\TAred{\hat{H}}_{\mathrm{c}}^{\TAred{\mathrm{R},}\Bar{i},\Bar{i}}(t)$ with eigenenergy
\begin{align}
    E_\mathrm{c}^{\bar{i},\bar{i}}(t)\TAred{/\hbar}:=
    -2g\alpha[\alpha_{1}(t)+\alpha_{2}]
    +K\alpha^4.
    \label{eq:eigenenrgy1}
\end{align}
In this article, we set $\chi$ and $|\alpha_{k}(t)|$ so small that
\begin{align}
    {\vphantom{\frac{\chi}{2}}}_{\mathrm{c}_k}\Braket{\mp\alpha_{k}(t)|
    \frac{\chi}{2}
    \hat{c}_{k}^{\dagger2}
    \hat{c}_{k}^2|
    \mp\alpha_{k}(t)}_{\mathrm{c}_k}
    =\frac{\chi}{2}\alpha_{k}(t)^4
\end{align}
can be neglected compared to $2g\alpha\alpha_{k}(t)$ in Eq.~\eqref{eq:eigenenrgy1}; the condition is
\begin{align}
    \chi|\alpha_{k}(t)|^3\ll g\alpha.
    \label{eq:condition1}
\end{align}
On the other hand, the tensor product of vacuum states $\ket{0,0}_{\mathrm{c}}$ is the eigenstate of Hamiltonian $\TAred{\hat{H}}_{\mathrm{c}}^{\TAred{\mathrm{R},}\Bar{i},\Bar{j}(\ne \bar{i})}(t)$ with eigenenergy
\begin{align}
    E_{\mathrm{c}}^{\bar{i},\bar{j}(\ne \bar{i})}(t)\TAred{/\hbar}= K\alpha^4.
    \label{eq:eigenenrgy2}
\end{align}
\TAred{When we set $\Delta_{1}(t)=-\Delta_{2}$, the first term in Eq.~\eqref{eq:eigenenrgy1} vanishes, and we have $E_\mathrm{c}^{\bar{i},\bar{i}}(t)=E_{\mathrm{c}}^{\bar{i},\bar{j}(\ne \bar{i})}(t)$, which means that four states, $\ket{\bar{0},\bar{0}}_{\mathrm{q}}\otimes\ket{-\alpha_{1}(t),-\alpha_{2}}_{\rm c}$,  $\ket{\bar{0},\bar{1}}_{\mathrm{q}}\otimes\ket{0,0}_{\mathrm{c}}$,  $\ket{\bar{1},\bar{0}}_{\mathrm{q}}\otimes\ket{0,0}_{\mathrm{c}}$, and $\ket{\bar{1},\bar{1}}_{\mathrm{q}}\otimes\ket{\alpha_{1}(t),\alpha_{2}}_{\rm c}$, are almost degenerate, so that we can suppress the residual $ZZ$ coupling.}

Suppose that the initial state of the system is represented as
\begin{align}
\ket{\Psi(0)}=\sum_{i,j=0}^1\beta_{\bar{i},\bar{j}}\ket{\bar{i},\bar{j}}_{\mathrm{q}}\otimes \ket{0,0}_{\mathrm{c}},
\label{eq:initial1}
\end{align} 
where $\beta_{\bar{i},\bar{j}}$ is a coefficient.
If we set $\Delta_{1}(0)=-\Delta_2$ and $|\Delta_2|\gg g\alpha$ so that the approximation $\ket{(-1)^{i+1}\alpha_{1}(0),(-1)^{i+1}\alpha_{2}}_{\rm c}\TAred{\approx}\ket{0,0}_{\mathrm{c}}$ is valid, $\ket{0,0}_{\mathrm{c}}$ is not only the eigenstate of $\TAred{\hat{H}}_{\mathrm{c}}^{\TAred{\mathrm{R},}\Bar{i},\Bar{j}(\ne \bar{i})}(0)$ but also the approximate eigenstate of $\TAred{\hat{H}}_{\mathrm{c}}^{\TAred{\mathrm{R},}\bar{i},\bar{i}}(0)$.
Hence, if we change $\Delta_{1}(t)$ slowly enough so that coupler $1$ evolves adiabatically, the system can evolve as
\begin{align}
\ket{\Psi(t)}\TAred{\approx}\sum_{i,j=0}^1
\mathcal{N}(t)\beta_{\bar{i},\bar{j}}e^{-i[\Theta_{\bar{i},\bar{j}}(t)+\theta(t)]}\ket{\bar{i},\bar{j}}_{\mathrm{q}}\otimes \ket{\psi(t)}_{\mathrm{c}}^{\bar{i},\bar{j}},
\label{Psi_3_14_23}
\end{align}
where
\begin{align}
\Theta_{\bar{i},\bar{j}}(t)&=
-\delta_{i,j}\int_0^{t}2g\alpha[\alpha_{1}(t')+\alpha_{2}]\,\mathrm{d}t'\nonumber\\
&=-4g^2\alpha^2\delta_{i,j}\int_0^{t}\left[
        \frac{1}{\Delta_{1}(t')}
        +\frac{1}{\Delta_{2}}
    \right]\,\mathrm{d}t',\\
\theta(t) &=
K\alpha^4 t,
\label{Theta_p_1_23_23} \\
\ket{\psi(t)}_{c}^{\bar{i},\bar{j}}
&=\delta_{i,j}\ket{(-1)^{i+1}\alpha_{1}(t),(-1)^{i+1}\alpha_{2}}_{\mathrm{c}} 
\notag \\
&\quad\mbox{}+(1-\delta_{i,j})\ket{0,0}_{\mathrm{c}},
\end{align}
and $\mathcal{N}(t)$ is for normalization.
After the gate time $t_f$, the detuning of coupler $1$
returns to the initial value so that $\ket{\psi(t_f)}_{\rm c}^{\bar{i},\bar{j}}\TAred{\approx}\ket{0,0}_{\rm c}$.
Defining $\Theta$ as
\begin{align}
\Theta=-4g^2\alpha^2\int_0^{t_f}\left[
        \frac{1}{\Delta_{1}(t)}
        +\frac{1}{\Delta_{2}}
    \right]\,\mathrm{d}t,
    \label{Theta_2_15_23}
\end{align}
we obtain
\begin{align}
\ket{\Psi(t_f)}&\approx\sum_{i,j=0}^1\mathcal{N}(t_f)\beta_{\bar{i},\bar{j}}\mathrm{e}^{-i[\Theta\delta_{i,j}+\theta(t_f)]}\ket{\bar{i},\bar{j}}_{\mathrm{q}}\otimes \ket{0,0}_{\mathrm{c}}
\notag \\
&=\mathrm{e}^{-i\theta(t_f)}\TAred{\hat{R}}_{zz}(\Theta)\ket{\Psi(0)}
=:\ket{\Psi^{\mathrm{ideal}}_{\Theta}},
\label{eq:final_state}
\end{align}
where $\theta(t_f)$ is the overall phase which is not of physical interest and
\begin{align}
    \TAred{\hat{R}}_{zz}(\Theta):=\mathcal{N}(t_f)\sum_{i,j=0}^1e^{-i\Theta\delta_{i,j}}\ket{\bar{i},\bar{j}}_{\mathrm{q}}\bra{\bar{i},\bar{j}}\otimes I_{\mathrm{c}}
\end{align}
is the $R_{zz}$ gate with rotation angle $\Theta$ on the Kerr-cat qubits with $I_{\mathrm{c}}$ being the identity operator on the couplers.
From Eq.~\eqref{eq:final_state}, we find that the $R_{zz}(\Theta)$ gate is approximately realized. 
Of note, we can eliminate \TAred{the} unwanted residual coupling by imposing $\Delta_{1}(t)=-\Delta_{2}$, which leads to $\Theta=0$ in Eq.~\eqref{Theta_2_15_23}.

\section{Numerical results}
\label{sec:gate}
We numerically examine the performance of our $ZZ$-coupling scheme.
\TAred{In this study, we take into account only single-photon loss of the KPOs and transmons as a source of decoherence, because it is dominant \cite[Supplementary Information, Sec.~II]{Puri2017}.
We assume that the single-photon loss rates of the KPOs and transmons are the same for simplicity.}
Time evolution of the system is simulated with the Gorini-Kossakowski-Sudarshan-Lindblad (GKSL)-type Markovian master equation \cite{doi:10.1063/1.522979,lindblad1976generators}:
\begin{alignat}{1}
    \frac{\mathrm{d}\hat{\rho}(t)}
    {\mathrm{d}t}
    &=-\frac{\mathrm{i}}{\hbar}
    [\hat{H}^{\mathrm{R}}(t),
    \hat{\rho}(t)]
    \notag \\
    &\mbox{}+\kappa\sum_{j=1,2}\left(
        \hat{a}_j\hat{\rho}(t)\hat{a}_j^{\dagger}
        -\frac{1}{2}
        \left\{
            \hat{a}_j^{\dagger}\hat{a}_j,\hat{\rho}(t)
        \right\}
    \right)
    \notag \\
    &\mbox{}+\kappa\sum_{k=1,2}\left(
        \hat{c}_{k}\hat{\rho}(t)\hat{c}_{k}^{\dagger}
        -\frac{1}{2}
        \left\{
            \hat{c}_{k}^{\dagger}\hat{c}_{k},\hat{\rho}(t)
        \right\}
    \right),
    \label{eq:gksl1}
\end{alignat}
where $\hat{\rho}(t)$ is the density operator of the system; $\kappa$ is the single-photon loss rate; $[\bullet,\circ]$ is the commutator; $\{\bullet,\circ\}$ is the anticommutator.
The used system parameters are listed in Table~\ref{table:parameter1}.
The initial state of the system is prepared as in Eq.~\eqref{eq:initial1} with $\beta_{\bar{i},\bar{j}}=[2(1+\mathrm{e}^{-2\alpha^2})]^{-1}$ $\forall \bar{i},\bar{j}\in\{0,1\}$.
That is, the four basis states of the Kerr-cat qubits are equally weighted.
The initial state can be rewritten as
\begin{align}
    \ket{\Psi(0)}&=\frac{\ket{\alpha}_{\mathrm{KPO}_1}+ \ket{-\alpha}_{\mathrm{KPO}_1}}{\sqrt{2(1+\mathrm{e}^{-2\alpha^2})}}
    \otimes
    \frac{\ket{\alpha}_{\mathrm{KPO}_2}+ \ket{-\alpha}_{\mathrm{KPO}_2}}{\sqrt{2(1+\mathrm{e}^{-2\alpha^2})}}
    \notag \\
    &\quad\otimes
    \ket{0,0}_{\mathrm{c}},
    \label{eq:initial2}
\end{align}
which shows that both of the KPOs are in the even-parity cat states.
\begin{table}
    \caption{The parameters of the system and detunings \TAred{used in Sec.~\ref{sec:gate}}. The used values of $K$ and $p$ correspond to $\alpha=2$.
    $\kappa/2\pi=0\,\mathrm{Hz}$ ($20$\,kHz) corresponds to the case without (with) decoherence.
    These parameters are comparable to those used in experiments \cite{Grimm2020_8_1,PhysRevA.105.023519,PhysRevX.9.021049_2019_6_7}.}
    \label{table:parameter1}
    \begin{center}
      \begin{tabular}{cccccc} \hline\hline
        $K/2\pi$ & $p/2\pi$ & $\chi/2\pi$ & $g/2\pi$ & $\kappa/2\pi$ & $\alpha_c^{\mathrm{min}}$ \\ \hline
        $20$\,MHz & $80$\,MHz & $10$\,MHz & $10$\,MHz & $0$\,Hz,\,$20$\,kHz & $0.04$ \\ \hline\hline
      \end{tabular}
    \end{center}
\end{table}

In numerical simulations, we used relevant energy eigenstates of isolated KPOs as basis to expand the states of the KPOs instead of their Fock states in order to reduce numerical resource, while Fock states are used for the couplers.
In a part of the simulations, we used Quantum Toolbox in Python (QuTiP) \cite{QuTiP2012,QuTiP2013}.

\subsection{Residual coupling}
\begin{figure}
    \centering
    \includegraphics[width=0.48\textwidth]{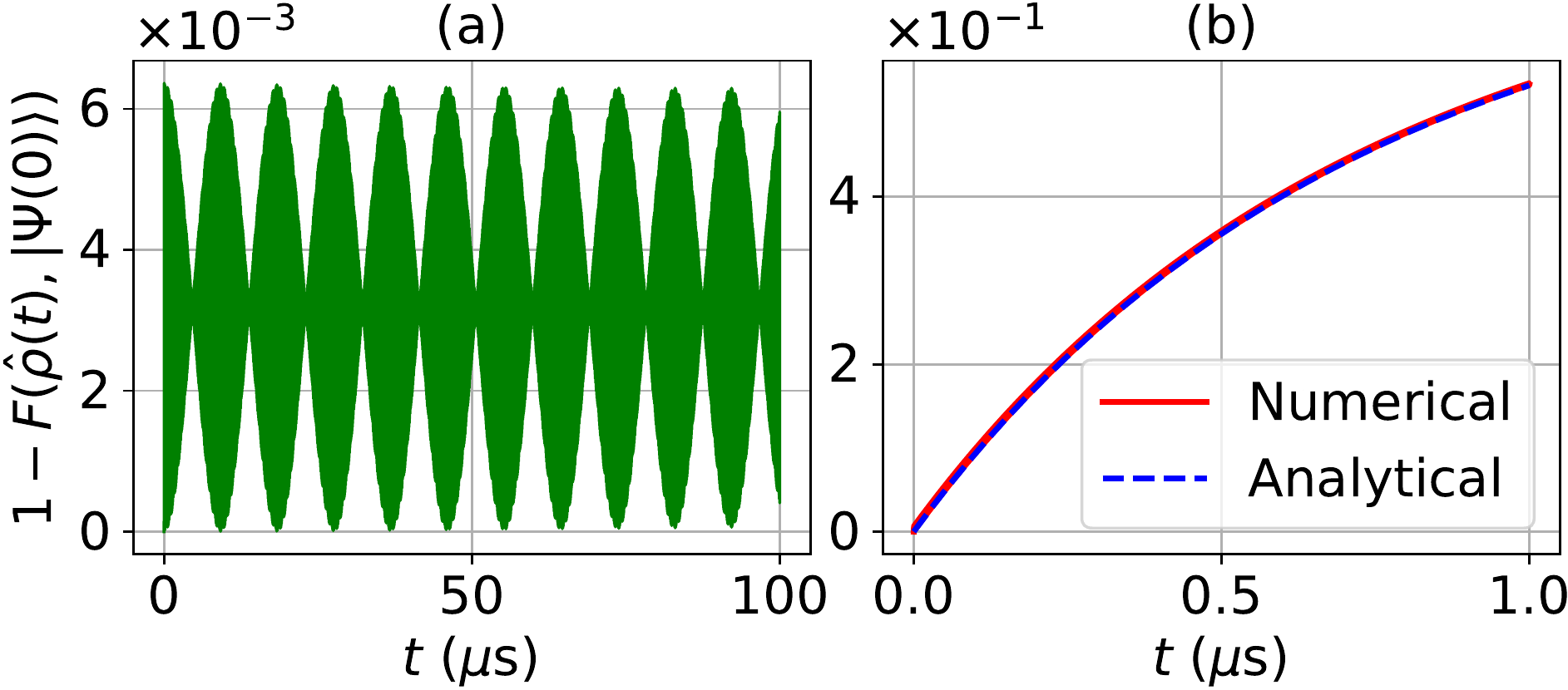}
    \caption{Infidelity $1-F(\hat{\rho}(t),\ket{\Psi(0)})$ when the effective $ZZ$ coupling should be switched {\scshape off} in the cases (a) without and (b) with decoherence.
    The red solid line labeled ``Numerical'' is calculated numerically with $\hat{\rho}(t)$ under master equation \eqref{eq:gksl1}.
    The blue dashed line labeled ``Analytical'' corresponds to Eq.~\eqref{eq:fidelity_dephased1}.
    The used parameters are shown in Table \ref{table:parameter1}.}
    \label{fig:off_infidelity1}
\end{figure}
We evaluate the performance of our scheme of \TAred{the} effective coupling when it is switched {\scshape off} with the infidelity defined as \cite[Sec.~9.2.2]{wilde_2017}
\begin{align}
    1-F(\hat{\rho}(t),\ket{\Psi(0)})
    =1-\langle\Psi(0)|\hat{\rho}(t)|\Psi(0)\rangle.
    \label{eq:fidelity1}
\end{align}
We set the detunings of the couplers as follows:
\begin{align}
    \Delta_{1}(t)
    =-\Delta_{2}
    =\frac{2g\alpha}
    {\alpha_c^{\mathrm{min}}}=2\pi\times1\,\mathrm{GHz}.
\end{align}
Figure \ref{fig:off_infidelity1} shows the infidelity $1-F(\hat{\rho}(t),\ket{\Psi(0)})$ for the cases \TAred{(a)} without and \TAred{(b)} with decoherence.
When the decoherence can be neglected, we do not see in  Fig.~\ref{fig:off_infidelity1}(a) the tendency of the infidelity to increase, which we regard as the sign that the residual coupling is suppressed.
We attribute the oscillation in the infidelity to the fact that the initial state is a superposition of different energy eigenstates of the system.

In contrast, when the decoherence cannot be neglected, we see in Fig.~\ref{fig:off_infidelity1}\TAred{(b)} the tendency of the infidelity to increase, so we must perform quantum error correction, which is left as our future work.
Let us explain this infidelity increase in terms of \TAred{Kerr-}cat\TAred{s} dephasing \TAred{due to single-photon loss of KPOs}.
\TAred{We assume that} the interaction between the KPOs and the couplers can be neglected, \TAred{which will be validated later.}
\TAred{Then,} the state of the system at time $t$ is written as \TAred{(see Appendix \ref{app:mapping})}
\begin{align}
    \hat{\rho}(t)&=\frac{\ket{\alpha}\bra{\alpha}
    +\ket{-\alpha}\bra{-\alpha}
    +\mathrm{e}^{-\gamma t}(\ket{\alpha}\bra{-\alpha}
    +\ket{-\alpha}\bra{\alpha})}{2(1+\mathrm{e}^{-2\alpha^2-\gamma t})}
    \notag \\
    &\otimes
    \frac{\ket{\alpha}\bra{\alpha}
    +\ket{-\alpha}\bra{-\alpha}
    +\mathrm{e}^{-\gamma t}(\ket{\alpha}\bra{-\alpha}
    +\ket{-\alpha}\bra{\alpha})}{2(1+\mathrm{e}^{-2\alpha^2-\gamma t})}
    \notag \\
    &\otimes\ket{0,0}_{\mathrm{c}}\bra{0,0},
    \label{eq:dephased_state1}
\end{align}
where the first (second) line in the right-hand side is the state of Kerr-cat qubit $1$ ($2$) \TAred{and $\gamma=2\kappa\alpha^2$ is the dephasing rate of the qubits}.
\TAred{T}he infidelity in Eq.~\eqref{eq:fidelity1} is calculated as
\begin{align}
    1-F(\hat{\rho}(t),\ket{\Psi(0)})
    =1-\frac{(1+\mathrm{e}^{-2\alpha^2})^2(1+\mathrm{e}^{-\gamma t})^2}{4(1+\mathrm{e}^{-2\alpha^2-\gamma t})^2}.
    \label{eq:fidelity_dephased1}
\end{align}
Since this agrees well with the numerical result as seen from Fig.~\ref{fig:off_infidelity1}\TAred{(b)}, the neglect of the interaction \TAred{between the KPOs and the couplers} is validated.
In other words, the residual coupling is suppressed in the presence of decoherence, too.

\subsection{Performance of the \texorpdfstring{$R_{zz}$}{Rzz} gate}
\label{subsec:Rzz_gate}
\begin{figure}
    \centering
    \includegraphics[width=0.4\textwidth]{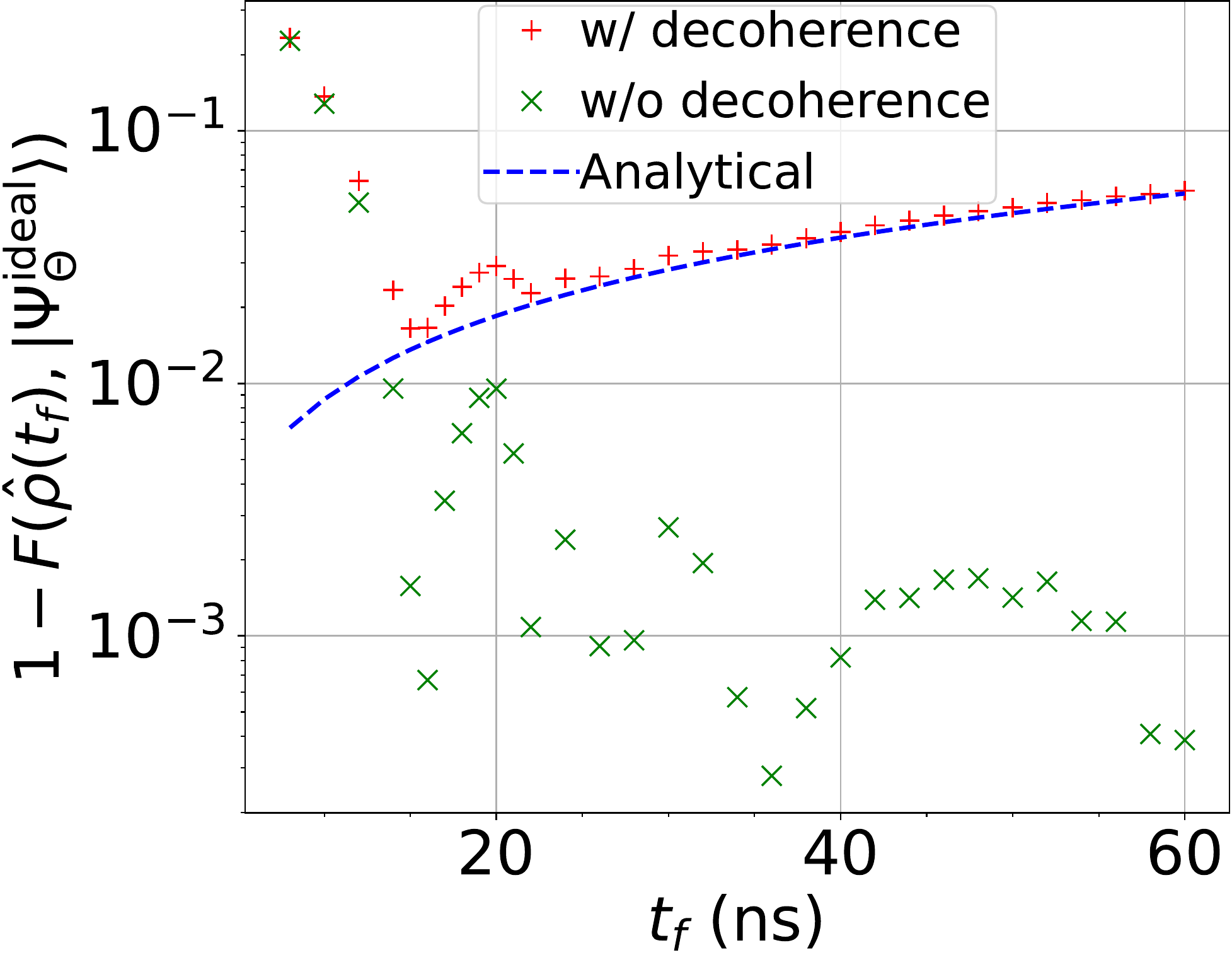}
    \caption{Infidelity of the $R_{zz}(-\pi/2)$ gate as a function of $t_f$, with and without decoherence.
    The blue dashed line labeled ``Analytical'' corresponds to Eq.~\eqref{eq:gate_infidelity_dehasing1}.
    The used parameters and $\alpha_{\rm c}^{\rm max}$  are listed in Table \ref{table:parameter1} and  Table~\ref{table:parameter2_alpha_c_max}, respectively.}
    \label{fig:gate_infidelity_1}
\end{figure}
The $R_{zz}$ gate for the Kerr-cat qubits can be realized by switching {\scshape on} their effective coupling.
We evaluate the performance of the $R_{zz}$ gate with the infidelity
defined as
\begin{align}
    1-F(\hat{\rho}(t_f),\ket{\Psi^{\mathrm{ideal}}_{\Theta}})
    =1-\langle\Psi^{\mathrm{ideal}}_{\Theta}|\hat{\rho}(t_f)|\Psi^{\mathrm{ideal}}_{\Theta}\rangle
    \label{fidelity2}
\end{align}
\TAred{under the detuning schedule in Appendix \ref{app:detuning_schedule}.}
We fix $\Theta=-\pi/2$ because $R_{zz}$ and $R_x$ ($X$-rotation) gates with rotation angle $-\pi/2$ and the $R_z$ ($Z$-rotation) gate with all rotation angles are sufficient to form a universal gate set \cite{PhysRevA.93.050301}.
Figure~\ref{fig:gate_infidelity_1} shows the
infidelity $1-F(\hat{\rho}(t_f),\ket{\Psi^{\mathrm{ideal}}_{\Theta}})$ as a function of $t_f$ in the cases with and without decoherence.
First, let us focus on the case without decoherence.
We find that the gate fidelity is over $99.9\%$ for $t_f=16$\,ns.
When $t_f\lesssim16$\,ns, the gate infidelity tends to decrease with the increase of $t_f$ due to the mitigation of nonadiabatic \TAred{errors}.
On the other hand, when $t_f\gtrsim16$\,ns, \TAred{the infidelity features an oscillating behavior.}
\TAred{This is because nonadiabatic errors are not necessarily mitigated monotonically with time.
They are related to the power spectral density of a gate-pulse form, and the graph of the density versus the gate time can show an oscillating behavior \cite{PhysRevA.90.022307}.
Therefore, the oscillating behavior in Fig.~\ref{fig:gate_infidelity_1} is not strange.}

Next, let us focus on the effect of decoherence.
When $t_f\lesssim12$\,ns, the effect is small, and the infidelities with and without decoherence are almost the same.
Otherwise, decoherence makes the infidelity larger; when $t_f\gtrsim 22$\,ns, the infidelity monotonically increases with $t_f$.
Let us explain this monotonic increase in a similar way with that in the last paragraph in the previous subsection.
From Eq.~\eqref{eq:final_state}, we have
\begin{align}
    \ket{\Psi^{\mathrm{ideal}}_{\Theta}}\bra{\Psi^{\mathrm{ideal}}_{\Theta}}
    &=\sum_{i,i',j,j'=0}^1\frac{\mathrm{e}^{\mathrm{i}\Theta(\delta_{i',j'}-\delta_{i,j})}}{4(1+2\cos{\Theta}\mathrm{e}^{-2\alpha^2}+\mathrm{e}^{-4\alpha^2})}
    \notag \\
    &\quad\times\ket{\bar{i}}_{\mathrm{q}_1}\bra{\bar{i'}}
    \otimes
    \ket{\bar{j}}_{\mathrm{q}_2}\bra{\bar{j'}}
    \otimes
    \ket{0,0}_{\mathrm{c}}\bra{0,0}.
\end{align}
This state does not include the effect of dephasing \TAred{of the Kerr-cat qubits}.
The counterpart with the effect is given by
\begin{align}
    &\quad\hat{\rho}^{\mathrm{dephased}}_{\Theta}(t_f)
    \notag \\
    &=\sum_{i,i',j,j'=0}^1
    \frac{\mathrm{e}^{\mathrm{i}\Theta(\delta_{i',j'}-\delta_{i,j})-\gamma t_f(2-\delta_{i,i'}-\delta_{j,j'})}}{4(1+2\cos{\Theta}\mathrm{e}^{-2\alpha^2-\gamma t_f}+\mathrm{e}^{-4\alpha^2-2\gamma t_f})}
    \notag \\
    &\quad\times\ket{\bar{i}}_{\mathrm{q}_1}\bra{\bar{i'}}
    \otimes
    \ket{\bar{j}}_{\mathrm{q}_2}\bra{\bar{j'}}
    \otimes
    \ket{0,0}_{\mathrm{c}}\bra{0,0}.    
\end{align}
The infidelity only due to the effect of dephasing is calculated as
\begin{align}
    &\quad1-F(\hat{\rho}^{\mathrm{dephased}}_{\Theta}(t_f),\ket{\Psi^{\mathrm{ideal}}_{\Theta}})
    \notag \\
    &=1-4^{-1}[(1+\mathrm{e}^{-\gamma t_f})^2(1+\mathrm{e}^{-4\alpha^2})
    \notag \\
    &\quad\times(1+4\cos{\Theta}\mathrm{e}^{-2\alpha^2}+\mathrm{e}^{-4\alpha^2})
    \notag \\
    &\quad\mbox{}+4\mathrm{e}^{-4\alpha^2}(1+2\cos{2\Theta}\mathrm{e}^{-\gamma t_f}+\mathrm{e}^{-2\gamma t_f})]
    \notag \\
    &\quad\times(1+2\cos{\Theta}\mathrm{e}^{-2\alpha^2}+\mathrm{e}^{-4\alpha^2})^{-1}
    \notag \\
    &\quad\times(1+2\cos{\Theta}\mathrm{e}^{-2\alpha^2-\gamma t_f}+\mathrm{e}^{-4\alpha^2-2\gamma t_f})^{-1}.
    \label{eq:gate_infidelity_dehasing1}
\end{align}
As this agrees well with the numerical result in Fig.~\ref{fig:gate_infidelity_1}, the infidelity increase for large $t_f$ is mostly due to dephasing.

\section{Conclusion}
\label{sec:conclusion}
We have developed a tunable $ZZ$-coupling scheme between two Kerr-cat qubits by using two transmon couplers.
The detuning of one of the couplers is tuned to control the coupling amplitude.
We have analytically presented and numerically confirmed that the residual coupling is eliminated by setting the detunings of the transmon couplers to values opposite to each other.
We have numerically achieved $R_{zz}(-\pi/2)$-gate fidelity over $99.9\%$ in the case of $16$-ns gate time and without decoherence.

As we have seen in Figs.~\ref{fig:off_infidelity1}\TAred{(b)} and \ref{fig:gate_infidelity_1}, the effect of \TAred{dephasing} is serious.
In order to preserve superposition states of Kerr-cat qubits, we must perform quantum error correction.
Combining our $ZZ$-coupling scheme with quantum error correction is left for future works.

\begin{acknowledgements}
The authors thank Yuichiro Matsuzaki, Toyofumi Ishikawa, and Hiroomi Chono for useful discussions.
This paper is based on results obtained from a project, JPNP16007, commissioned by the New Energy and Industrial Technology Development Organization (NEDO), Japan.
\end{acknowledgements}

\appendix
\section{\TAred{Circuit model I}}
\label{app:Circuit model}
\begin{figure}
    \centering
    \includegraphics[width=0.47\textwidth]{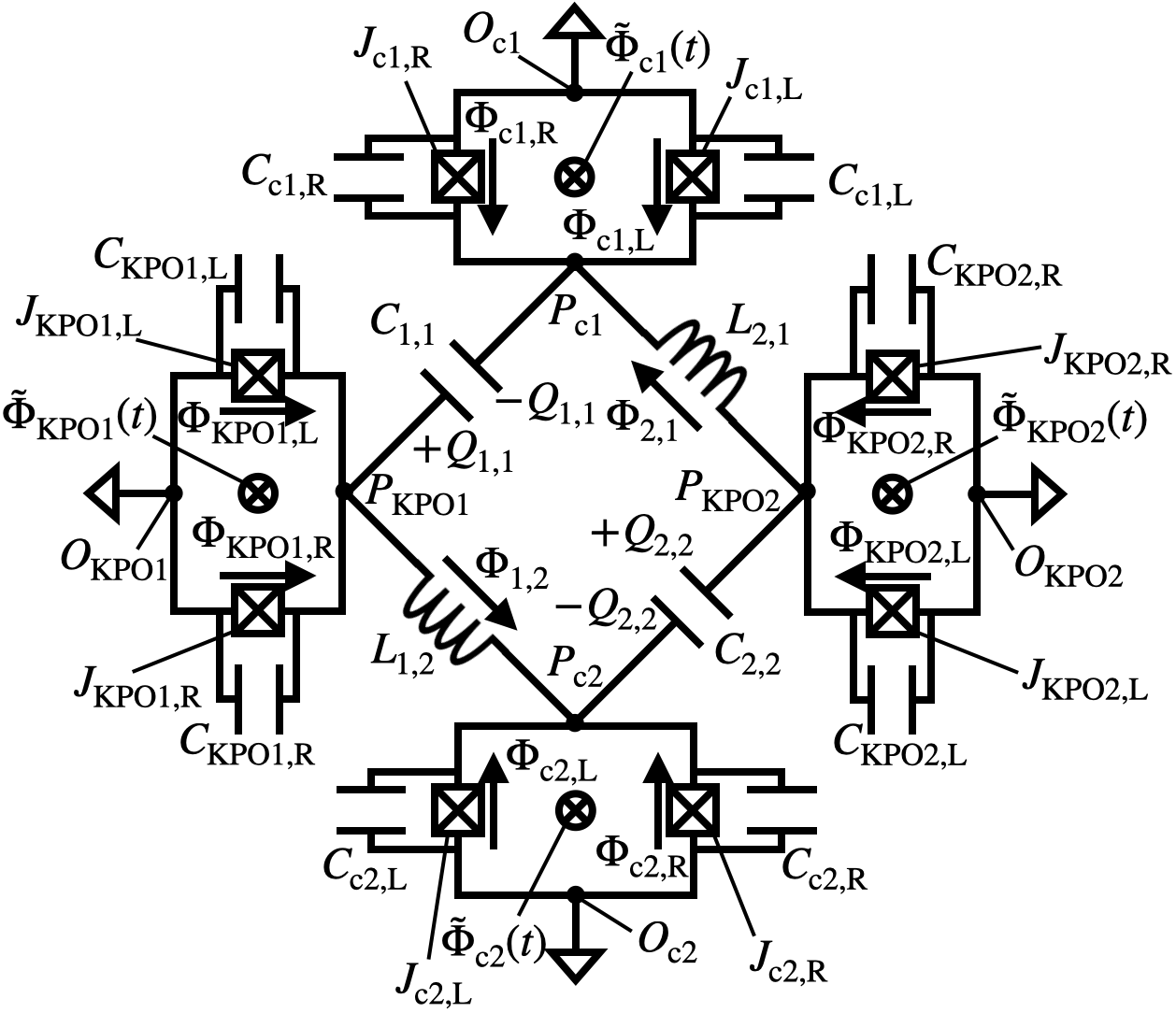}
    \caption{A circuit diagram to describe Hamiltonian \eqref{eq:H_R_t_main1}.}
    \label{fig:system2}
\end{figure}
In this Appendix, we give a circuit example to realize our scheme in the main text; see Fig.~\ref{fig:system2}.
We derive its effective Hamiltonian \eqref{eq:H_R_t_main1} using canonical quantization \cite{PRXQuantum.2.040204} and standard methods \cite{doi:10.7566/JPSJ.88.061015_2019_3_1,Masuda2021,PhysRevB.84.174521}.
A similar circuit in Fig.~\ref{fig:system3} is discussed in Appendix \ref{app:Circuit model2}.
Each subsystem, $\lambda\in\{\mathrm{KPO}1$, $\mathrm{KPO}2$, $\mathrm{c}1 (\mathrm{coupler}1)$, $\mathrm{c}2 (\mathrm{coupler}2)\}$, comprises a symmetric dc SQUID which has two Josephson junctions $J_{\lambda,\mathrm{L}}$ and $J_{\lambda,\mathrm{R}}$ of Josephson energy $E^{\lambda}_J$ and two shunting capacitors $C_{\lambda,\mathrm{L}}$ and $C_{\lambda,\mathrm{R}}$;
we set  $C_{\lambda,\mathrm{L}}=C_{\lambda,\mathrm{R}}=:C_{\lambda}/2$.
KPO $j$ and coupler $k$ ($j,k\in\{1,2\}$) are coupled via a capacitor $C_{j,k}$ for $j=k$ and via a linear inductor $L_{j,k}$ for $j\neq k$.
The branch flux variables across the Josephson junctions  $J_{\lambda,\mathrm{L}}$ and $J_{\lambda,\mathrm{R}}$ are denoted by $\Phi_{\lambda,\mathrm{L}}$ and $\Phi_{\lambda,\mathrm{R}}$, respectively.
Then, the branch charge variables across the capacitors $C_{\lambda,\mathrm{L}}$ and $C_{\lambda,\mathrm{R}}$ are given by $C_{\lambda,\mathrm{L}}\dot{\Phi}_{\lambda,\mathrm{L}}$ and $C_{\lambda,\mathrm{R}}\dot{\Phi}_{\lambda,\mathrm{R}}$, respectively.
The branch flux variable across the linear inductor $L_{j,k}$ and the branch charge variable across the capacitor $C_{j,k}$ are denoted by $\Phi_{j,k}$ and $Q_{j,k}$, respectively.
Each dc SQUID is threaded by a time-dependent magnetic flux $\tilde{\Phi}_{\lambda}(t)$ directed into the paper.
We assume that the magnetic flux $\tilde{\Phi}_{\lambda}(t)$ is concentrated at the center of the SQUID of $\lambda$ in a symmetric distribution with regard to horizontal and perpendicular axes through the center \cite[Sec.~12]{2022arXiv220101945N}.
The circuit of $\lambda$ is symmetric with respect to the straight line through $P_{\lambda}$ and $O_{\lambda}$.
The point $O_{\lambda}$ is connected to the ground.
Following normal procedure in electrodynamics \cite{Riwar2022,McDonald2018voltage}, we obtain
\begin{gather}
    \Phi_{\lambda,\mathrm{L}}
    =\Phi_{\lambda}+\frac{\tilde{\Phi}_{\lambda}(t)}{2},
    \quad
    \Phi_{\lambda,\mathrm{R}}
    =\Phi_{\lambda}-\frac{\tilde{\Phi}_{\lambda}(t)}{2}, \\
    \Phi_{1,2}=\Phi_{\mathrm{c}2}-\Phi_{\mathrm{KPO}1},
    \quad
    \Phi_{2,1}=\Phi_{\mathrm{c}1}-\Phi_{\mathrm{KPO}2}, \\
    \frac{Q_{1,1}}{C_{1,1}}
    =\dot{\Phi}_{\mathrm{KPO}1}-\dot{\Phi}_{\mathrm{c}1},
    \quad
    \frac{Q_{2,2}}{C_{2,2}}
    =\dot{\Phi}_{\mathrm{KPO}2}-\dot{\Phi}_{\mathrm{c}2},
\end{gather}
where
\begin{align}
    \Phi_{\lambda}:=\frac{\Phi_{\lambda,\mathrm{L}}+\Phi_{\lambda,\mathrm{R}}}{2}.
    \label{eq:Phi_lambda_def}
\end{align}

-Let us derive Hamiltonian \eqref{eq:H_1}, which describes Fig.~\ref{fig:system2}, by canonical quantization \cite{PRXQuantum.2.040204}.
The kinetic energy of the circuit is written as
\begin{align}
    T&=\sum_{\lambda}\left[\frac{(C_{\lambda,\mathrm{L}}\dot{\Phi}_{\lambda,\mathrm{L}})^2}{2C_{\lambda,\mathrm{L}}}
    +\frac{(C_{\lambda,\mathrm{R}}\dot{\Phi}_{\lambda,\mathrm{R}})^2}{2C_{\lambda,\mathrm{R}}}\right]
    +\frac{Q_{1,1}^2}{2C_{1,1}}
    +\frac{Q_{2,2}^2}{2C_{2,2}}
    \notag \\
    &=\frac{1}{2}\dot{\boldsymbol{\Phi}}^{\mathrm{T}}M\dot{\boldsymbol{\Phi}}    =\frac{\Phi_0^2}{2}\dot{\boldsymbol{\varphi}}^{\mathrm{T}}M\dot{\boldsymbol{\varphi}},
\end{align}
where
\begin{gather}
    \boldsymbol{\Phi}:=\begin{pmatrix}
        \Phi_{\mathrm{KPO}1} \\
        \Phi_{\mathrm{KPO}2} \\
        \Phi_{\mathrm{c}1} \\
        \Phi_{\mathrm{c}2}        
    \end{pmatrix},\,    
    \boldsymbol{\varphi}:=\frac{\boldsymbol{\Phi}}{\Phi_0}=\begin{pmatrix}
        \varphi_{\mathrm{KPO}1} \\
        \varphi_{\mathrm{KPO}2} \\
        \varphi_{\mathrm{c}1} \\
        \varphi_{\mathrm{c}2}        
    \end{pmatrix},
    \, \varphi_{\lambda}:=\frac{\Phi_{\lambda}}{\Phi_0},   
\end{gather}
$\Phi_0=\hbar/(2e)$ is the reduced flux quantum [$h/(2e)$ is the flux quantum],
\begin{align}
    M=\begin{pmatrix}
        C+\tilde{C} & 0 & -\tilde{C} & 0 \\
        0 & C+\tilde{C} & 0 & -\tilde{C} \\
        -\tilde{C} & 0 & C+\tilde{C} & 0 \\
        0 & -\tilde{C} & 0 & C+\tilde{C}
    \end{pmatrix}
\end{align}
(we have set $C_{\mathrm{KPO}1}=C_{\mathrm{KPO}2}=C_{\mathrm{c}1}=C_{\mathrm{c}2}=:C$ and $C_{1,1}=C_{2,2}=:\tilde{C}$ for brevity), and we have neglected the terms $\sum_{\lambda}C\left[\dot{\tilde{\Phi}}_{\lambda}(t)\right]^2/8$, which do not affect the dynamics of the system.
The potential energy of the circuit is written as
\begin{align}
    U(t)&=-\sum_{\lambda}\Biggl[
    E_J^{\lambda}\cos{\left(\frac{\Phi_{\lambda,\mathrm{L}}}{\Phi_0}\right)}+E_J^{\lambda}\cos{\left(\frac{\Phi_{\lambda,\mathrm{R}}}{\Phi_0}\right)}\Biggr]
    \notag \\
    &\quad\mbox{}+\frac{\Phi_{1,2}^2}{L_{2,1}}+\frac{\Phi_{2,1}^2}{L_{2,1}}
    \notag \\
    &=-\sum_{\lambda}2E_J^{\lambda}\cos{\left(\frac{\phi_{\lambda}(t)}{2}\right)}
    \cos{\varphi_{\lambda}}
    \notag \\
    &\quad\mbox{}+\frac{E_L}{2}(\varphi_{\mathrm{c}2}-\varphi_{\mathrm{KPO}1})^2+\frac{E_L}{2}(\varphi_{\mathrm{c}1}-\varphi_{\mathrm{KPO}2})^2,
    \label{eq:Potential2}
\end{align}
where $\phi_{\lambda}(t)=\tilde{\Phi}_{\lambda}(t)/\Phi_0$, we have set $L_{1,2}=L_{2,1}=:L$, and $E_L=\Phi_0^2/L$.
The Lagrangian of the system in the laboratory frame is written as
\begin{align}
    \mathcal{L}(t)=T-U(t).
\end{align}

We define $n_{\lambda}$ as the conjugate momentum to $\hbar\varphi_{\lambda}$:
\begin{align}
    \boldsymbol{n}=\frac{1}{\hbar}\frac{\partial \mathcal{L}}{\partial\dot{\boldsymbol{\varphi}}}=\frac{\Phi_0^2}{\hbar}M\dot{\boldsymbol{\varphi}}.
\end{align}
The classical Hamiltonian in the laboratory frame is obtained from the following Legendre transformation:
\begin{align}
    H(t)&=\hbar\dot{\boldsymbol{\varphi}}\cdot\boldsymbol{n}-\mathcal{L}(t)=2e^2\boldsymbol{n}^{\mathrm{T}}M^{-1}\boldsymbol{n}+U(t),
    \label{eq:classical_H}
\end{align}
where
\begin{gather}
    M^{-1}=\frac{1}{C}\begin{pmatrix}
        u & 0 & v & 0 \\
        0 & u & 0 & v \\
        v & 0 & u & 0 \\
        0 & v & 0 & u
    \end{pmatrix}
\end{gather}
with
\begin{gather}
    u=\frac{1+x}{1+2x},
    \quad
    v=\frac{x}{1+2x},
    \quad
    x=\frac{\tilde{C}}{C}.
\end{gather}
Then, Hamiltonian \eqref{eq:classical_H} is rewritten as
\begin{align}
    H(t)&=4E_C \Biggl[u\sum_{\lambda}n_{\lambda}^2
    +2v\sum_{j=1}^2n_{\mathrm{KPO}j}n_{\mathrm{c}j}
    \Biggr]+U(t)
    \label{eq:classical_H2}
\end{align}
with $E_C=e^2/(2C)$.
By quantization, $n_{\lambda}\to\hat{n}_{\lambda}$ and $\varphi_{\lambda}\to\hat{\varphi}_{\lambda}$ with $[\hat{\varphi}_{\lambda},\hat{n}_{\lambda'}]=\mathrm{i}\delta_{\lambda,\lambda'}$, we obtain the following quantum Hamiltonian in the laboratory frame:
\begin{align}
    \hat{H}(t)&=\sum_{\lambda}\hat{H}_{\lambda}(t)+\hat{H}_I,
    \label{eq:H_1}\\
    \hat{H}_{\lambda}(t)&=4uE_C\hat{n}_{\lambda}^2
    -2E_J^{\lambda}\cos{\left(\frac{\phi_{\lambda}(t)}{2}\right)}\cos{\hat{\varphi}_{\lambda}}
    +\frac{E_L}{2}\hat{\varphi}_{\lambda}^2,
    \label{eq:H_lambda1} \\
    \hat{H}_I&=8vE_C\sum_{j=1}^2\hat{n}_{\mathrm{KPO}j}\hat{n}_{\mathrm{c}j}
    \notag \\
    &\quad\mbox{}-E_L\hat{\varphi}_{\mathrm{KPO}1}\hat{\varphi}_{\mathrm{c}2}-E_L\hat{\varphi}_{\mathrm{KPO}2}\hat{\varphi}_{\mathrm{c}1}.
    \label{eq:H_I1}
\end{align}

Then, let us derive Hamiltonian \eqref{eq:H_R_t_main1} by conventional methods \cite{doi:10.7566/JPSJ.88.061015_2019_3_1,Masuda2021,PhysRevB.84.174521}. 
In Eq.~\eqref{eq:H_lambda1}, we assume $uE_C\ll E_J^{\lambda}\cos{[\phi_{\lambda}(t)/2]}$ $\forall t$.
Then, expanding $\cos{\hat{\varphi}_{\lambda}}$ to the fourth order in $\hat{\varphi}_{\lambda}$ is a good approximation:
\begin{align}
    \cos{\hat{\varphi}_{\lambda}}\approx1-\frac{1}{2}\hat{\varphi}_{\lambda}^2+\frac{1}{24}\hat{\varphi}_{\lambda}^4.
    \label{eq:cos_fourth}
\end{align}
Let us decompose $\phi_{\lambda}(t)$ into dc and ac parts \cite[Sec.~4.1]{doi:10.7566/JPSJ.88.061015_2019_3_1}:
\begin{gather}
    \phi_{\lambda}(t)=\phi_{\lambda}^{\mathrm{dc}}(t)+\phi_{\lambda}^{\mathrm{ac}}(t), \\
    \phi_{\lambda}^{\mathrm{ac}}(t)
    =-2\pi\epsilon_{p,\lambda}\cos{(\omega_{p,\lambda}t)}.
\end{gather}
We set $|\epsilon_{p,\lambda}|\ll1$, so that the following approximation may be good \cite[Sec.~4.1]{doi:10.7566/JPSJ.88.061015_2019_3_1}:
\begin{align}
    \hat{H}_{\lambda}(t)&\approx4uE_C\hat{n}_{\lambda}^2
    +\frac{\tilde{E}_J^{\lambda,\mathrm{dc}}(t)+E_L}{2}\hat{\varphi}_{\lambda}^2
    +\frac{\tilde{E}_J^{\lambda,\mathrm{ac}}(t)}{2}\hat{\varphi}_{\lambda}^2
    \notag \\
    &\quad\mbox{}-\frac{\tilde{E}_J^{\lambda,\mathrm{dc}}(t)}{24}\hat{\varphi}_{\lambda}^4,
    \label{eq:H_lambda2}
\end{align}
where
\begin{align}
    \tilde{E}_J^{\lambda,\mathrm{dc}}(t)&:=2E_J^{\lambda}\cos{\left(\frac{\phi_{\lambda}^{\mathrm{dc}}(t)}{2}\right)},
    \label{eq:EJ_tilde_dc1} \\
    \tilde{E}_J^{\lambda,\mathrm{ac}}(t)&:=2\pi\epsilon_{p,\lambda}E_J^{\lambda}\sin{\left(\frac{\phi_{\lambda}^{\mathrm{dc}}(t)}{2}\right)}\cos{(\omega_{p,\lambda}t)}.
    \label{eq:EJ_tilde_ac1}
\end{align}

We introduce the annihilation and creation operators, $\hat{a}_{\lambda}$ and $\hat{a}_{\lambda}^{\dagger}$, through \cite[Sec.~II]{PhysRevB.84.174521}
\begin{align}
    \hat{\varphi}_{\lambda}&=
    \left(\frac{2uE_C}{\tilde{E}_J^{\lambda,\mathrm{dc}}(0)+E_L}
    \right)^{1/4}
    \left(\hat{a}_{\lambda}+\hat{a}_{\lambda}^{\dagger}\right),
    \label{eq:hat_phi1}\\
    \hat{n}_{\lambda}&=
    -\frac{\mathrm{i}}{2}
    \left(\frac{\tilde{E}_J^{\lambda,\mathrm{dc}}(0)+E_L}{2uE_C}
    \right)^{1/4}
    \left(\hat{a}_{\lambda}-\hat{a}_{\lambda}^{\dagger}\right).
    \label{eq:hat_n1}
\end{align}
Then Eq.~\eqref{eq:H_lambda2} is rewritten as \cite{Masuda2021}
\begin{align}
    \hat{H}_{\lambda}(t)/\hbar&\approx
    \omega_{\lambda}^{(0)}(t)\hat{a}_{\lambda}^{\dag}\hat{a}_{\lambda}
    +\frac{[\omega_{\lambda}^{(0)}(t)-\omega_{\lambda}^{(0)}(0)]}{2}\left(\hat{a}_{\lambda}^{\dag2}+\hat{a}_{\lambda}^{2}\right)
    \notag \\
    &\quad\mbox{}-\frac{ K_{\lambda}(t)}{12}\left(\hat{a}_{\lambda}+\hat{a}_{\lambda}^{\dagger}\right)^4
    \notag \\
    &\quad\mbox{}+ p_{\lambda}(t)\left(\hat{a}_{\lambda}+\hat{a}_{\lambda}^{\dagger}\right)^2\cos{(\omega_{p,\lambda}t)},
\end{align}
where
\begin{align}
    \omega_{\lambda}^{(0)}(t)&:=\frac{\sqrt{2uE_C[\tilde{E}_J^{\lambda,\mathrm{dc}}(0)+E_L]}}{\hbar}
    \notag \\
    &\quad\mbox{}\times\left(\frac{\tilde{E}_J^{\lambda,\mathrm{dc}}(t)+E_L}{\tilde{E}_J^{\lambda,\mathrm{dc}}(0)+E_L}+1\right),
    \label{eq:omega_c_lambda_0_t1} \\
    K_{\lambda}(t)&:=\frac{uE_C}{\hbar}\frac{\tilde{E}_J^{\lambda,\mathrm{dc}}(t)}{\tilde{E}_J^{\lambda,\mathrm{dc}}(0)+E_L},
    \label{eq:K_lambda_t1} \\
    p_{\lambda}(t)&:=\frac{\pi\epsilon_{p,\lambda}E_J^{\lambda}}{\hbar}\sqrt{\frac{2uE_C}{\tilde{E}_J^{\lambda,\mathrm{dc}}(0)+E_L}}\sin{\left(\frac{\phi_{\lambda}^{\mathrm{dc}}(t)}{2}\right)},
    \label{eq:p_lambda_t1}
\end{align}
and we have omitted a c-valued term.
Inserting Eqs.~\eqref{eq:hat_phi1}
 and \eqref{eq:hat_n1} into Eq.~\eqref{eq:H_I1} yields 
\begin{align}
    \hat{H}_I/\hbar&=-\sum_{j=1}^2g_{j,j}
    \left(\hat{a}_{\mathrm{KPO}j}-\hat{a}_{\mathrm{KPO}j}^{\dagger}\right)
    \left(\hat{a}_{\mathrm{c}j}-\hat{a}_{\mathrm{c}j}^{\dagger}\right)
    \notag \\
    &\quad\mbox{}+g_{1,2}\left(\hat{a}_{\mathrm{KPO}1}+\hat{a}_{\mathrm{KPO}1}^{\dagger}\right)
    \left(\hat{a}_{\mathrm{c}2}+\hat{a}_{\mathrm{c}2}^{\dagger}\right)
    \notag \\
    &\quad\mbox{}+g_{2,1}\left(\hat{a}_{\mathrm{KPO}2}+\hat{a}_{\mathrm{KPO}2}^{\dagger}\right)
    \left(\hat{a}_{\mathrm{c}1}+\hat{a}_{\mathrm{c}1}^{\dagger}\right),
\end{align}
where
\begin{align}
    g_{j,j}&:=\frac{v}{\hbar}\sqrt{\frac{2E_C}{u}}
    \left[\tilde{E}_J^{\mathrm{KPO}j,\mathrm{dc}}(0)+E_L\right]^{1/4}
    \notag \\
    &\quad\times\left[\tilde{E}_J^{\mathrm{c}j,\mathrm{dc}}(0)+E_L\right]^{1/4},
    \label{eq:g_jj_1} \\
    g_{1,2}&=-\frac{E_L\sqrt{2uE_C}}{\hbar\left[\tilde{E}_J^{\mathrm{KPO}1,\mathrm{dc}}(0)+E_L\right]^{1/4}\left[\tilde{E}_J^{\mathrm{c}2,\mathrm{dc}}(0)+E_L\right]^{1/4}},
    \label{g_12_1}\\
    g_{2,1}&=-\frac{E_L\sqrt{2uE_C}}{\hbar\left[\tilde{E}_J^{\mathrm{KPO}2,\mathrm{dc}}(0)+E_L\right]^{1/4}\left[\tilde{E}_J^{\mathrm{c}1,\mathrm{dc}}(0)+E_L\right]^{1/4}}.
    \label{g_21_1}
\end{align}

We set $\omega_{p,\mathrm{KPO}1}=\omega_{p,\mathrm{KPO}2}=\omega_{p,\mathrm{c}1}=\omega_{p,\mathrm{c}2}=:\omega_p$.
Let us move on to a rotating frame at frequency $\omega_p/2$ and neglect rapidly oscillating terms (rotating-wave approximation), so that the Hamiltonian there may be \cite{Masuda2021}
\begin{align}
    \hat{H}^{\mathrm{R}}(t)&=
    \hat{R}(t)\hat{H}(t)\hat{R}^{\dagger}(t)+\mathrm{i}\hbar\left(\frac{\mathrm{d}\hat{R}(t)}{\mathrm{d}t}\right)\hat{R}^{\dagger}(t)
    \notag \\
    &\approx\hat{H}^{\mathrm{R}}_{\lambda}(t)+\hat{H}^{\mathrm{R}}_I(t),
\end{align}
where
\begin{align}
    \hat{R}(t)&=\exp\left[
    \mathrm{i}\frac{\omega_p}{2}t\sum_{\lambda}\hat{a}_{\lambda}^{\dag}\hat{a}_{\lambda}\right],
    \\
    \hat{H}_{\lambda}^{\mathrm{R}}(t)/\hbar&=\Delta_{\lambda}(t)\hat{a}_{\lambda}^{\dagger}\hat{a}_{\lambda}-\frac{K_{\lambda}(t)}{2}\hat{a}_{\lambda}^{\dagger2}\hat{a}_{\lambda}^2
    \notag \\
    &\quad\mbox{}+\frac{p_{\lambda}(t)}{2}\left(\hat{a}_{\lambda}^{\dagger2}+\hat{a}_{\lambda}^2\right),
    \label{eq:H_lambda_R2} \\
    \hat{H}_I^{\mathrm{R}}/\hbar&=\sum_{j,k=1}^2g_{j,k}
    \left(\hat{a}_{\mathrm{KPO}j}\hat{a}_{\mathrm{c}k}^{\dagger}+\hat{a}_{\mathrm{KPO}j}^{\dagger}\hat{a}_{\mathrm{c}k}\right),
    \label{eq:HIR2}
\end{align}
where $\Delta_{\lambda}(t):=\omega_{\lambda}(t)-\omega_p/2$ and $\omega_{\lambda}(t):=\omega_{\lambda}^{(0)}(t)-K_{\lambda}(t)$.

We set $E_J^{\mathrm{KPO}1}=E_J^{\mathrm{KPO2}}=:E_J^{\mathrm{KPO}}$, $\epsilon_{p,\mathrm{KPO}1}=\epsilon_{p,\mathrm{KPO}2}=:\epsilon_p$, and $\phi_{\mathrm{KPO}1}^{\mathrm{dc}}(t)=\phi_{\mathrm{KPO}1}^{\mathrm{dc}}(0)=\phi_{\mathrm{KPO}2}^{\mathrm{dc}}(0)=:\phi_{\mathrm{KPO}}^{\mathrm{dc}}=\phi_{\mathrm{KPO}2}^{\mathrm{dc}}(t)$ $\forall t$.
From Eqs.~\eqref{eq:EJ_tilde_dc1}, \eqref{eq:omega_c_lambda_0_t1}, \eqref{eq:K_lambda_t1}, and \eqref{eq:p_lambda_t1}, we obtain $\tilde{E}_{J}^{\mathrm{KPO}1,\mathrm{dc}}(t)=\tilde{E}_{J}^{\mathrm{KPO}1,\mathrm{dc}}(0)=\tilde{E}_{J}^{\mathrm{KPO}2,\mathrm{dc}}(0)=:\tilde{E}_{J}^{\mathrm{KPO},\mathrm{dc}}=\tilde{E}_{J}^{\mathrm{KPO}2,\mathrm{dc}}(t)$ $\forall t$,
$\omega_{\mathrm{KPO}1}^{(0)}(t)=\omega_{\mathrm{KPO}2}^{(0)}(t)=2\sqrt{2uE_C(\tilde{E}_J^{\mathrm{KPO,dc}}+E_L)}/\hbar=:\omega_{\mathrm{KPO}}^{(0)}$ $\forall t$, $K_{\mathrm{KPO}1}(t)=K_{\mathrm{KPO}2}(t)=uE_C\tilde{E}_J^{\mathrm{KPO,dc}}/\hbar(\tilde{E}_J^{\mathrm{KPO,dc}}+E_L)=:K$ $\forall t$, $p_{\mathrm{KPO}1}(t)=p_{\mathrm{KPO}2}(t)=\pi\epsilon_pE_J^{\mathrm{KPO}}\sqrt{2uE_C/(\tilde{E}_J^{\mathrm{KPO,dc}}+E_L)}\sin(\phi_{\mathrm{KPO}}^{\mathrm{dc}}/2)/\hbar=:p$ $\forall t$.
We also have $\omega_{\mathrm{KPO}1}(t)=\omega_{\mathrm{KPO}2}(t)=\omega_{\mathrm{KPO}}^{(0)}-K=:\omega_{\mathrm{KPO}}$ $\forall t$.
We set $\omega_p=2\omega_{\mathrm{KPO}}$ so that we may satisfy $\Delta_{\mathrm{KPO}1}(t)=\Delta_{\mathrm{KPO}2}(t)=0$ $\forall t$.
We define $\hat{a}_{\mathrm{KPO}j}=:\hat{a}_j$ for $j=1,2$.
From Eq.~\eqref{eq:H_lambda_R2}, we obtain
\begin{align}
    \hat{H}_{\mathrm{KPO}j}^{\mathrm{R}}/\hbar=-\frac{K}{2}\hat{a}_j^{\dag2}\hat{a}_j^2+\frac{p}{2}\left(\hat{a}_j^{\dag2}+\hat{a}_j^2\right).
    \label{eq:H_KPOj_R1}
\end{align}

We set $\epsilon_{p,\mathrm{c}1}=\epsilon_{p,\mathrm{c}2}=0$, which gives $p_{\mathrm{c}1}(t)=p_{\mathrm{c}2}(t)=0$ $\forall t$.
We define $\Delta_{\mathrm{c}k}(t)=:\Delta_{k}(t)$, $K_{\mathrm{c}k}(t)=:\chi_k(t)$, and $\hat{a}_{\mathrm{c}k}=:\hat{c}_k$ for $k=1,2$.
From \eqref{eq:H_lambda_R2} and \eqref{eq:HIR2}, we have
\begin{align}
    \hat{H}_{\mathrm{c}k}^{\mathrm{R}}(t)/\hbar&=
    -\frac{\chi_k(t)}{2}\hat{c}_k^{\dagger2}\hat{c}_k^2+\Delta_k(t)\hat{c}_k^{\dagger}\hat{c}_k,
    \label{H_ck_R_t1} \\
    \hat{H}_I^{\mathrm{R}}/\hbar&=\sum_{j,k=1}^2g_{j,k}\left(\hat{a}_j\hat{c}_k^{\dagger}+\hat{a}_j^{\dagger}\hat{c}_k\right).
\end{align}
Hence we arrive at Eq.~\eqref{eq:H_R_t_main1}.

\section{\TAred{Circuit model II}}
\label{app:Circuit model2}
\begin{figure}
    \centering
    \includegraphics[width=0.47\textwidth]{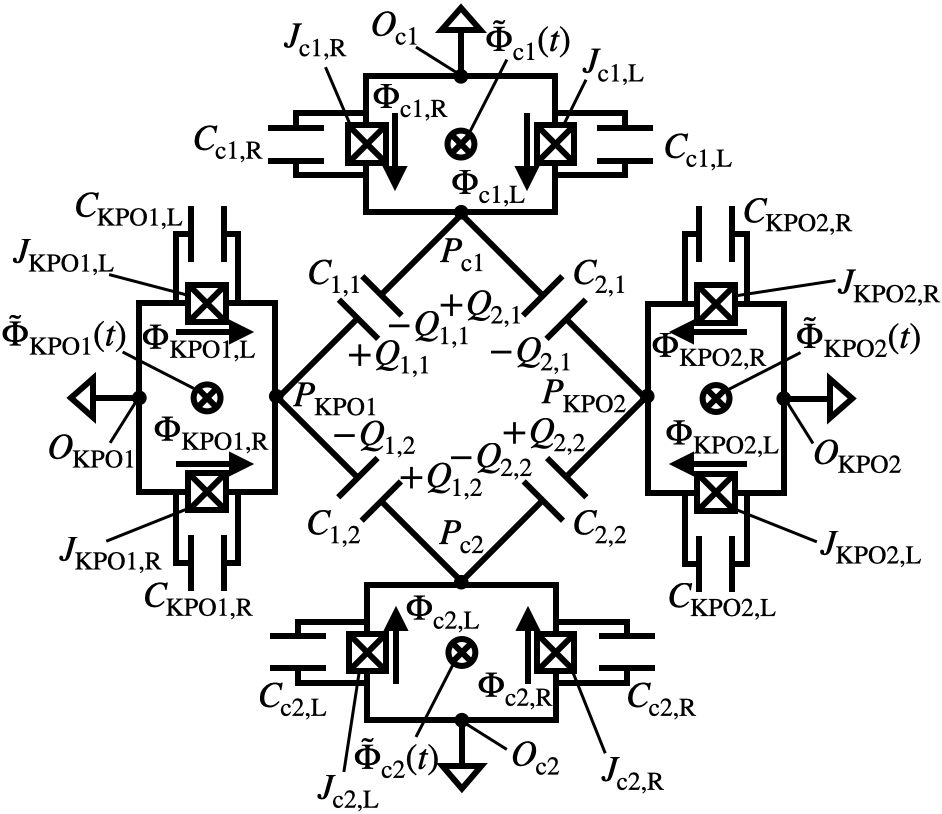}
    \caption{A circuit diagram to describe Hamiltonian \eqref{eq:H_R_t_main2}.
    The difference between this circuit and the circuit in Fig.~\ref{fig:system2} is $\{C_{1,2},C_{2,1}\}$ and $\{L_{1,2},L_{2,1}\}$.}
    \label{fig:system3}
\end{figure}
In this Appendix, we explain that we can suppress the residual $ZZ$ coupling between Kerr-cat qubits in the case of a circuit which replaces $L_{1,2}$ and $L_{2,1}$ in Fig.~\ref{fig:system2} with $C_{1,2}(=\tilde{C})$ and $C_{2,1}(=\tilde{C})$; see Fig.~\ref{fig:system3}.
Following the similar procedure to that in Appendix \ref{app:Circuit model}, the quantum Hamiltonian of the system in a rotating frame at frequency $\omega_p/2$ is obtained, in rotating-wave approximation, as
\begin{align}
    \hat{\tilde{H}}^{\mathrm{R}}(t)&=\sum_{j=1}^2\hat{\tilde{H}}_{\mathrm{KPO}j}^{\mathrm{R}}
    +\sum_{k=1}^2\hat{\tilde{H}}_{\mathrm{c}k}^{\mathrm{R}}(t)+\hat{\tilde{H}}_I^{\mathrm{R}},
    \label{eq:H_R_t_main2} \\
    \hat{\tilde{H}}_{\mathrm{KPO}j}^{\mathrm{R}}/\hbar&=-\frac{\tilde{K}}{2}\hat{a}_j^{\dag2}\hat{a}_j^2+\frac{\tilde{p}}{2}\left(\hat{a}_j^{\dag2}+\hat{a}_j^2\right),
    \label{eq:H_KPOj_R_main2} \\
    \hat{\tilde{H}}_{\mathrm{c}k}^{\mathrm{R}}(t)/\hbar&=
    -\frac{\tilde{\chi}_k(t)}{2}\hat{c}_k^{\dagger2}\hat{c}_k^2+\tilde{\Delta}_k(t)\hat{c}_k^{\dagger}\hat{c}_k,
    \label{eq:H_ck_R_t_main2} \\
    \hat{\tilde{H}}_I^{\mathrm{R}}/\hbar&=\sum_{j,k=1}^2g_k\left(\hat{a}_j\hat{c}_k^{\dagger}+\hat{a}_j^{\dagger}\hat{c}_k\right)
    \notag \\
    &\quad\mbox{}+g_{\mathrm{KPO}}\left(\hat{a}_1\hat{a}_2^{\dagger}+\hat{a}_1^{\dagger}\hat{a}_2\right)+g_{\mathrm{c}}\left(\hat{c}_1\hat{c}_2^{\dagger}+\hat{c}_1^{\dagger}\hat{c}_2\right),
    \label{eq:H_I_R_main2}
\end{align}
where
\begin{align}
    \tilde{K}&=yE_C/\hbar, \\
    \tilde{p}&=\frac{\pi\epsilon_pE_J^{\mathrm{KPO}}}{\hbar}\sqrt{\frac{2yE_C}{\tilde{E}_J^{\mathrm{KPO,dc}}}}\sin\left(\frac{\phi_{\mathrm{KPO}}^{\mathrm{dc}}}{2}\right), \\
    \tilde{\chi}_k(t)&=\frac{yE_C}{\hbar}\frac{\tilde{E}_J^{\mathrm{c}k,\mathrm{dc}}(t)}{\tilde{E}_J^{\mathrm{c}k,\mathrm{dc}}(0)}, \\
    \tilde{\Delta}_k(t)&=\frac{\sqrt{2yE_C\tilde{E}_J^{\mathrm{c}k,\mathrm{dc}}(0)}}{\hbar}\left(\frac{\tilde{E}_J^{\mathrm{c}k,\mathrm{dc}}(t)}{\tilde{E}_J^{\mathrm{c}k,\mathrm{dc}}(0)}+1\right)-\tilde{\chi}_k(t)
    \notag \\
    &\quad\mbox{}-\frac{2\sqrt{2yE_C\tilde{E}_J^{\mathrm{KPO},\mathrm{dc}}}}{\hbar}+\tilde{K}, \\
    g_k&=\frac{z}{\hbar}\sqrt{\frac{E_C}{y}}
    \left[\tilde{E}_J^{\mathrm{KPO},\mathrm{dc}}\tilde{E}_J^{\mathrm{c}k,\mathrm{dc}}(0)\right]^{1/4}, \\
    g_{\mathrm{KPO}}&:=\frac{w}{\hbar}\sqrt{\frac{E_C\tilde{E}_J^{\mathrm{KPO},\mathrm{dc}}}{y}}, \\
    g_{\mathrm{c}}&:=\frac{w}{\hbar}\sqrt{\frac{E_C}{y}}
    \left[\tilde{E}_J^{\mathrm{c}1,\mathrm{dc}}(0)\tilde{E}_J^{\mathrm{c}2,\mathrm{dc}}(0)\right]^{1/4}, \\
    y&=\frac{1+4x+2x^2}{1+6x+8x^2},
    \quad
    z=\frac{x}{1+4x}, \\
    w&=\frac{2x^2}{1+6x+8x^2},
    \quad
    x=\frac{\tilde{C}}{C}.
\end{align}

\begin{figure}
    \centering
    \includegraphics[width=0.47\textwidth]{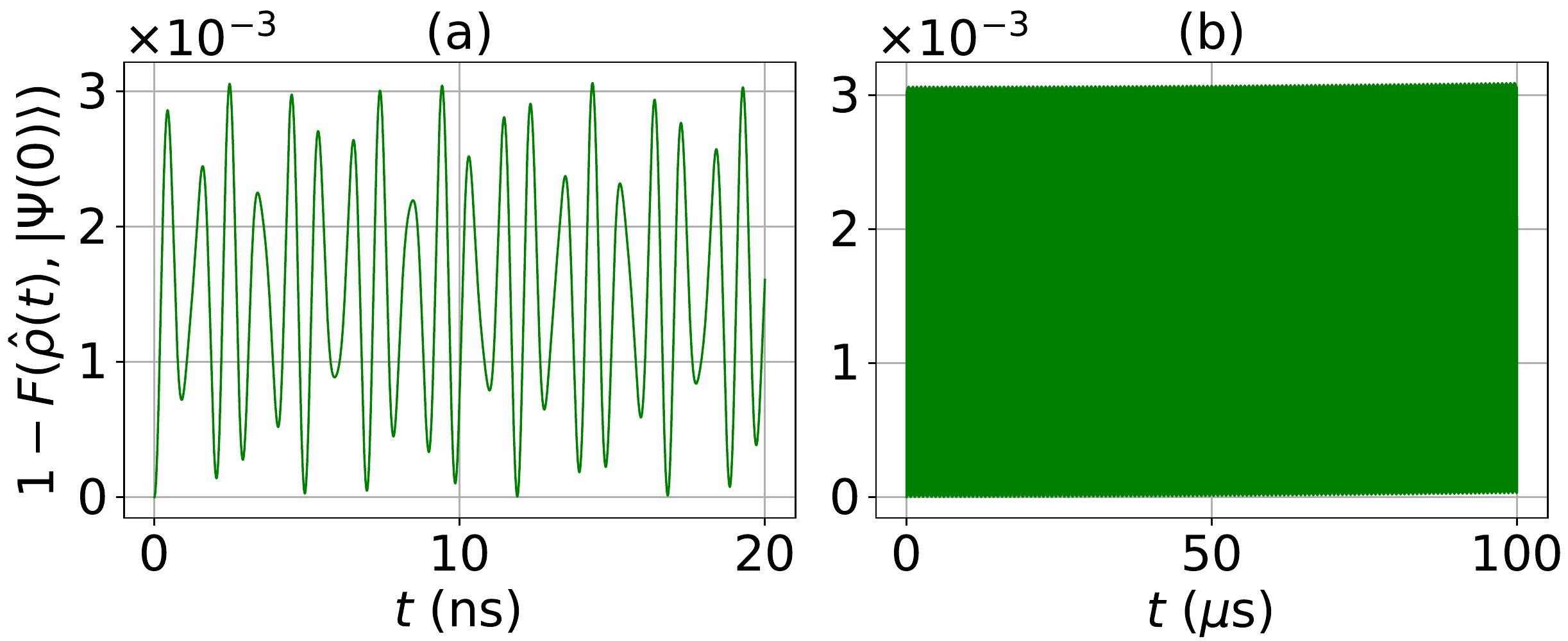}
    \caption{Infidelity $1-F(\hat{\rho}(t),\ket{\Psi(0)})$ when the effective $ZZ$ coupling should be switched {\scshape off} in the case without decoherence:
    (a) $0\,\mathrm{ns}\leq t\leq20\,$ns;
    (b) $0\,\mu\mathrm{s}\leq t\leq100\,\mu$s.
    The used parameters are shown in Table~\ref{table:parameter3}.}
    \label{fig:off_infidelity2}
\end{figure}
\begin{table}
    \caption{Parameter values used in Fig.~\ref{fig:off_infidelity2}.
    The bold values are design values, from which the other values (only the first three digits are shown) are calculated.
    $\kappa/(2\pi)=0\,\mathrm{Hz}$ corresponds to the case without decoherence.}
    \label{table:parameter3}
    \begin{center}
      \begin{tabular}{cccc}
      \hline\hline
      $C$ (pF) & $\boldsymbol{1.1}$ & $E_J^{\mathrm{c}1}/h$ (GHz) & $\boldsymbol{660}$ \\
      $\tilde{C}$ (fF) & $\boldsymbol{2}$ & $E_J^{\mathrm{c}2}/h$ (GHz) & $\boldsymbol{444.69}$ \\
      $E_C/h$ (MHz) & $17.6$ & $\phi_{\mathrm{c}1}^{\mathrm{dc}}(t)$ & $\boldsymbol{0}$ \\
      $x$ & $1.82\times10^{-3}$ & $\phi_{\mathrm{c}2}^{\mathrm{dc}}$ & $\boldsymbol{0}$ \\
      $y$ & $0.996$ & $\tilde{E}_J^{\mathrm{c}1,\mathrm{dc}}(t)/h$ (THz) & $1.32$ \\
      $z$ & $1.81\times10^{-3}$ & $\tilde{E}_J^{\mathrm{c}2,\mathrm{dc}}/h$ (GHz) & $889$ \\
      $w$ & $6.54\times10^{-6}$ & $\tilde{\chi}_{\mathrm{c}1}(t)/(2\pi)$ (MHz) & $17.5$ \\
      $E_J^{\mathrm{KPO}}/h$ (GHz) & $\boldsymbol{800}$ & $\tilde{\chi}_{\mathrm{c}2}/(2\pi)$ (MHz) & $17.5$ \\
      $\phi_{\mathrm{KPO}}^{\mathrm{dc}}$ & $\boldsymbol{\pi/2}$ & $\tilde{\Delta}_1(t)/(2\pi)$ (GHz) & $1.01$ \\
      $\tilde{E}_J^{\mathrm{KPO,dc}}/h$ (THz) & $1.13$ & $\tilde{\Delta}_2/(2\pi)$ (GHz) & $-1.43$ \\
      $\epsilon_p$ & $\boldsymbol{7\times10^{-3}}$ & $g_{1}/(2\pi)$ (MHz) & $8.39$ \\
      $\tilde{K}/(2\pi)$ (MHz) & $17.5$ & $g_{2}/(2\pi)$ (MHz) & $7.60$ \\
      $\tilde{p}/(2\pi)$ (MHz) & $69.3$ & $g_{\mathrm{KPO}}/(2\pi)$ (kHz) & $29.2$ \\
      $\tilde{\alpha}$ & $1.99$ & $g_{\mathrm{c}}/(2\pi)$ (kHz) & $28.6$ \\
      $\kappa/(2\pi)$ (Hz) & $\boldsymbol{0}$ \\
      \hline\hline
      \end{tabular}
    \end{center}
\end{table}

Although Eq.~\eqref{eq:H_I_R_main2} contains the extra terms compared to Eq.~\eqref{eq:H_I_R_main1}, we can suppress the residual $ZZ$ coupling in essentially the same way as that explained in Sec.~\ref{sec:coupling}.
The counterparts to Eqs.~\eqref{eq:H_eff_c_same_sign_1}--\eqref{eq:eigenenrgy1}, and \eqref{eq:eigenenrgy2} are
\begin{align}
    \hat{\tilde{H}}_{\mathrm{c}}^{\mathrm{R},\Bar{i},\Bar{i}}(t)/\hbar
    &=\tilde{\Delta}_1(t)
    \left(\hat{c}_1^{\dag}+\frac{g_{\mathrm{c}}}{\tilde{\Delta}_1(t)}\hat{c}_2^{\dag}+\frac{(-1)^i2g_1\tilde{\alpha}}{\tilde{\Delta}_1(t)}\right)
    \notag \\
    &\quad\mbox{}\cdot\left(\hat{c}_1+\frac{g_{\mathrm{c}}}{\tilde{\Delta}_1(t)}\hat{c}_2+\frac{(-1)^i2g_1\tilde{\alpha}}{\tilde{\Delta}_1(t)}\right)
    \notag \\
    &\quad\mbox{}+\left(
    \tilde{\Delta}_2-\frac{g_{\mathrm{c}}^2}{\tilde{\Delta}_1(t)}
    \right)
    \notag \\
    &\quad\mbox{}\cdot\left(
    \hat{c}_2^{\dag}+\frac{(-1)^i2\tilde{\alpha}[g_2\tilde{\Delta}_1(t)-g_1g_{\mathrm{c}}]}{\tilde{\Delta}_1(t)\tilde{\Delta}_2-g_{\mathrm{c}}^2}
    \right)
    \notag \\
    &\quad\mbox{}\cdot\left(
    \hat{c}_2+\frac{(-1)^i2\tilde{\alpha}[g_2\tilde{\Delta}_1(t)-g_1g_{\mathrm{c}}]}{\tilde{\Delta}_1(t)\tilde{\Delta}_2-g_{\mathrm{c}}^2}
    \right)
    \notag \\
    &\quad\mbox{}+\sum_{k=1}^2
    \left(-\frac{\tilde{\chi}_k(t)}{2}
    \hat{c}_{k}^{\dagger2}
    \hat{c}_{k}^2\right)
    \notag \\
    &\quad\mbox{}+2g_{\mathrm{KPO}}\tilde{\alpha}^2+\tilde{K}\tilde{\alpha}^4    
    \notag \\
    &\quad\mbox{}-\frac{4\tilde{\alpha}^2[g_1^2\tilde{\Delta}_2+g_2^2\tilde{\Delta}_1(t)-2g_1g_2g_{\mathrm{c}}]}{\tilde{\Delta}_1(t)\tilde{\Delta}_2-g_{\mathrm{c}}^2}, \\
    \hat{\tilde{H}}_{\mathrm{c}}^{\mathrm{R},\Bar{i},\Bar{j}(\ne\Bar{i})}(t)/\hbar
    &=\sum_{k=1}^2\left[-\frac{\tilde{\chi}_k(t)}{2} \hat{c}_{k}^{\dagger2}
    \hat{c}_{k}^2
    +\tilde{\Delta}_k(t)\hat{c}_k^{\dagger}\hat{c}_k\right] 
    \notag \\
    &\quad\mbox{}+g_{\mathrm{c}}\left(\hat{c}_1\hat{c}_2^{\dagger}+\hat{c}_1^{\dagger}\hat{c}_2\right)-2g_{\mathrm{KPO}}\tilde{\alpha}^2+\tilde{K}\tilde{\alpha}^4, \\
    \tilde{E}_{\mathrm{c}}^{\mathrm{R},\Bar{i},\Bar{i}}(t)/\hbar
    &=2g_{\mathrm{KPO}}\tilde{\alpha}^2+\tilde{K}\tilde{\alpha}^4    
    \notag \\
    &\quad\mbox{}-\frac{4\tilde{\alpha}^2[g_1^2\tilde{\Delta}_2+g_2^2\tilde{\Delta}_1(t)-2g_1g_2g_{\mathrm{c}}]}{\tilde{\Delta}_1(t)\tilde{\Delta}_2-g_{\mathrm{c}}^2}, \\
    \tilde{E}_{\mathrm{c}}^{\mathrm{R},\Bar{i},\Bar{j}(\ne\Bar{i})}(t)/\hbar&=-2g_{\mathrm{KPO}}\tilde{\alpha}^2+\tilde{K}\tilde{\alpha}^4,
\end{align}
where $\tilde{\alpha}:=\sqrt{\tilde{p}/\tilde{K}}$.
When we set
\begin{align}
    \tilde{\Delta}_1(t)=\frac{g_1^2\tilde{\Delta}_2+g_{\mathrm{KPO}}g_{\mathrm{c}}^2-2g_1g_2g_{\mathrm{c}}}{g_{\mathrm{KPO}}\tilde{\Delta}_2-g_2^2},
    \label{eq:detuning_allC1}
\end{align}
we have $\tilde{E}_{\mathrm{c}}^{\mathrm{R},\Bar{i},\Bar{i}}(t)=\tilde{E}_{\mathrm{c}}^{\mathrm{R},\Bar{i},\Bar{j}(\ne \Bar{i})}(t)$,
so that we can suppress the residual $ZZ$ coupling.
We show in Fig.~\ref{fig:off_infidelity2} the infidelity $1-F(\hat{\rho}(t),\ket{\Psi(0)})$ for the case without decoherence under parameter values in Table~\ref{table:parameter3}, which almost satisfy Eq.~\eqref{eq:detuning_allC1}.
The infidelity oscillates as in Fig.~\ref{fig:off_infidelity2}(a) and hardly tends to increase with time as in Fig.~\ref{fig:off_infidelity2}(b).

\section{The states of the qubits during the control of \texorpdfstring{$\Delta_1(t)$}{Delta1}}
\label{app:qubits_states}
In order to verify that the states of the Kerr-cat qubits remain $\ket{\bar{i},\bar{j}}_{\rm q}$ during the control of $\Delta_1(t)$ \TAred{in Sec.~\ref{sec:coupling}}, we consider the effective drive Hamiltonians of the KPOs conditioned by the state of the couplers, which is either $\ket{(-1)^{i+1}\alpha_{1}(t),(-1)^{i+1}\alpha_{2}}_{\rm c}$ or $\ket{0,0}_{\rm c}$.
We define the effective drive Hamiltonians corresponding to $\ket{(-1)^{i+1}\alpha_{1}(t),(-1)^{i+1}\alpha_{2}}_{\rm c}$ and $\ket{0,0}_{\rm c}$ as
\begin{widetext}
\begin{align}
    \hat{H}_{I,\mathrm{KPO}}^{\TAred{\mathrm{R},}\bar{i},\bar{i}}(t)\TAred{/\hbar}:&={\vphantom{\Ket{(-1)^{i+1}}}}_{\mathrm{c}}\Braket{(-1)^{i+1}\alpha_{1}(t),(-1)^{i+1}\alpha_{2}|\hat{H}_I^{\TAred{\mathrm{R}}}\TAred{/\hbar}|(-1)^{i+1}\alpha_{1}(t),(-1)^{i+1}\alpha_{2}}_{\mathrm{c}}
    \notag \\
    &=(-1)^{i+1}g[\alpha_{1}(t)+\alpha_{2}]\sum_{j=1}^2\left(\hat{a}_j^{\dag}+\hat{a}_j
    \right)
\end{align}
\end{widetext}
and
\begin{align}
    \hat{H}_{I,\mathrm{KPO}}^{\TAred{\mathrm{R},}\bar{i},\bar{j}(\neq\bar{i})}:&={\vphantom{\Braket{0,0|\hat{H}_I\TAred{^{\mathrm{R}}}|0,0}}}_{\mathrm{c}}\Braket{0,0|\hat{H}_I|0,0}_{\mathrm{c}}=0,
\end{align}
respectively.
Since $|\alpha_{1}(t)|$ and $|\alpha_2|$ are set to be small in this study, not only $ \hat{H}_{I,\mathrm{KPO}}^{\bar{i},\bar{j}(\neq\bar{i})}$ but also $\hat{H}_{I,\mathrm{KPO}}^{\bar{i},\bar{i}}(t)$ does not deviate the state of the Kerr-cat qubits from $\ket{\bar{i},\bar{j}}_{\rm q}$ very much.

\section{Derivation of Eq.~\eqref{eq:dephased_state1}}
\label{app:mapping}
If the interaction Hamiltonian $\hat{H}_I^{\mathrm{R}}$ can be neglected, master equation \eqref{eq:gksl1} and the initial state \eqref{eq:initial2} give
\begin{align}
    \hat{\rho}(t)=\hat{\rho}_{\mathrm{KPO}1}(t)\otimes\hat{\rho}_{\mathrm{KPO}2}(t)\otimes\hat{\rho}_{\mathrm{c}1}(t)\otimes\hat{\rho}_{\mathrm{c}2}(t)
\end{align}
with
\begin{align}
    &\frac{\mathrm{d}\hat{\rho}_{\mathrm{KPO}j}(t)}
    {\mathrm{d}t}
    =-\frac{\mathrm{i}}{\hbar}
    [\hat{H}^{\mathrm{R}}_{\mathrm{KPO}j},
    \hat{\rho}_{\mathrm{KPO}j}(t)]
    \notag \\
    &\qquad\mbox{}+\kappa\bigg(
        \hat{a}_j\hat{\rho}_{\mathrm{KPO}j}(t)\hat{a}_j^{\dagger}
    -\frac{1}{2}
        \left\{
            \hat{a}_j^{\dagger}\hat{a}_j,\hat{\rho}_{\mathrm{KPO}j}(t)
        \right\}
    \bigg),
    \label{eq:me_KPO1}
    \\
    &\frac{\mathrm{d}\hat{\rho}_{\mathrm{c}k}(t)}
    {\mathrm{d}t}
    =-\frac{\mathrm{i}}{\hbar}
    [\hat{H}^{\mathrm{R}}_{\mathrm{c}k},
    \hat{\rho}_{\mathrm{c}k}(t)]
    \notag \\
    &\qquad\mbox{}+\kappa\bigg(
        \hat{c}_k\hat{\rho}_{\mathrm{c}k}(t)\hat{c}_k^{\dagger}
    -\frac{1}{2}
        \left\{
            \hat{c}_k^{\dagger}\hat{c}_k,\hat{\rho}_{\mathrm{c}k}(t)
        \right\}
    \bigg).
    \label{eq:me_coupler1}
\end{align}
Since the initial state of coupler $k$, $\ket{0}_{\mathrm{c}k}$, is the steady state of master equation \eqref{eq:me_coupler1}, we have
\begin{align}
    \hat{\rho}_{\mathrm{c}k}(t)=\ket{0}_{\mathrm{c}k}\bra{0}.
\end{align}

We map master equation \eqref{eq:me_KPO1} of KPO $j$
onto that of Kerr-cat qubit $j$.
For the time being, we use
\begin{align}
    \ket{C_{\pm}}_{\mathrm{q}j}:=\mathcal{N}_{\pm}(\ket{\alpha}_{\mathrm{KPO}j}\pm\ket{-\alpha}_{\mathrm{KPO}j})
    \label{eq:basis2}
\end{align}
with $\mathcal{N}_{\pm}:=[2(1\pm\mathrm{e}^{-2\alpha^2})]^{-1/2}$ as two basis states of qubit $j$.
The annihilation operator is represented in the qubit subspace as \cite[Sec.~III]{PhysRevX.9.041009_2019_10_9}
\begin{align}
    \hat{a}_j=r\alpha\ket{C_{-}}_{\mathrm{q}j}\bra{C_{+}}+r^{-1}\alpha\ket{C_{+}}_{\mathrm{q}j}\bra{C_{-}}
\end{align}
with $r=\mathcal{N}_{+}/\mathcal{N}_{-}$.
Its matrix representation is
\begin{align}
    \hat{a}_j =\alpha\begin{pmatrix}
        0 & r^{-1} \\
        r & 0
    \end{pmatrix}.
    \label{eq:aj_matrix1}
\end{align}
Using this, the Hamiltonian of KPO $j$ is mapped onto that of qubit $j$ as
\begin{align}
    \hat{H}^{\mathrm{R}}_{\mathrm{q}j}=\frac{K\alpha^4}{2}I,
    \label{eq:Hami_matrix1}
\end{align}
where $I$ is the identity matrix with dimension two.
The density matrix of qubit $j$ in the Bloch sphere representation \cite[Sec.~III B]{doi:10.1063/1.5089550} is written as
\begin{align}
    \hat{\rho}_{\mathrm{q}j}(t)&=\frac{I+\vec{b}_j(t)\cdot\vec{\sigma}}{2}
    \notag \\
    &=\frac{1}{2}\begin{pmatrix}
        1+b_{j,z}(t) & b_{j,x}(t)-\mathrm{i}b_{j,y}(t) \\
        b_{j,x}(t)+\mathrm{i}b_{j,y}(t) & 1-b_{j,z}(t)
    \end{pmatrix},
    \label{eq:rho_matrix1}
\end{align}
where $\vec{b}_j(t)=[b_{j,x}(t),\,b_{j,y}(t),\,b_{j,z}(t)]^{\mathrm{T}}$ with $|\vec{b}_j(t)|\leq1$ is the Bloch vector and $\vec{\sigma}=[\sigma_x,\,\sigma_y,\,\sigma_z]$ is a vector of Pauli matrices.
Using Eqs.~\eqref{eq:me_KPO1}, \eqref{eq:aj_matrix1}--\eqref{eq:rho_matrix1}, the master equation of qubit $j$ is written as
\begin{align}
    \frac{\mathrm{d}\hat{\rho}_{\mathrm{q}j}(t)}
    {\mathrm{d}t}
    &=-\frac{\mathrm{i}}{\hbar}
    [\hat{H}^{\mathrm{R}}_{\mathrm{q}j},
    \hat{\rho}_{\mathrm{q}j}(t)]
    \notag \\
    &\quad\mbox{}+\kappa\Bigg(
        \hat{a}_j\hat{\rho}_{\mathrm{q}j}(t)\hat{a}_j^{\dagger}
    -\frac{1}{2}
        \left\{
            \hat{a}_j^{\dagger}\hat{a}_j,\hat{\rho}_{\mathrm{q}j}(t)
        \right\}
    \Bigg)
\end{align}
with
\begin{align}
    \frac{\mathrm{d}\hat{\rho}_{\mathrm{q}j}(t)}
    {\mathrm{d}t}
    =\frac{1}{2}\begin{pmatrix}
        \dot{b}_{j,z}(t) & \dot{b}_{j,x}(t)-\mathrm{i}\dot{b}_{j,y}(t) \\
        \dot{b}_{j,x}(t)+\mathrm{i}\dot{b}_{j,y}(t) & -\dot{b}_{j,z}(t)
    \end{pmatrix},
\end{align}
\begin{widetext}
\begin{align}
    &-\frac{\mathrm{i}}{\hbar}
    [\hat{H}^{\mathrm{R}}_{\mathrm{q}j},
    \hat{\rho}_{\mathrm{q}j}(t)]
    +\kappa\Bigg(
        \hat{a}_j\hat{\rho}_{\mathrm{q}j}(t)\hat{a}_j^{\dagger}
    -\frac{1}{2}
        \left\{
            \hat{a}_j^{\dagger}\hat{a}_j,\hat{\rho}_{\mathrm{q}j}(t)
        \right\}
    \Bigg)
    \notag \\
    &\qquad=\frac{\kappa\alpha^2}{2}\begin{pmatrix}
        -(r^{-2}+r^2)b_{j,z}(t)+r^{-2}-r^2
        & b_{j,x}(t)+\mathrm{i}b_{j,y}(t)-\frac{r^{-2}-r^2}{2}[b_{j,x}(t)-\mathrm{i}b_{j,y}] \\
        b_{j,x}(t)-\mathrm{i}b_{j,y}(t)-\frac{r^{-2}-r^2}{2}[b_{j,x}(t)+\mathrm{i}b_{j,y}]
        & (r^{-2}+r^2)b_{j,z}(t)-r^{-2}+r^2
    \end{pmatrix}.
\end{align}
\end{widetext}
The solution is
\begin{align}
    b_{j,x}(t)&=\exp{\bigg[
    -\kappa\alpha^2\bigg(\frac{r^{-2}+r^2}{2}-1\bigg)t
    \bigg]}b_{j,x}(0),
    \\
    b_{j,y}(t)&=\exp{\bigg[
    -\kappa\alpha^2\bigg(\frac{r^{-2}+r^2}{2}+1\bigg)t
    \bigg]}b_{j,y}(0),
    \\
    b_{j,z}(t)&=\exp{[-\kappa\alpha^2(r^{-2}+r^2)t]}\bigg[b_{j,z}(0)-\frac{r^{-2}-r^2}{r^{-2}+r^2}\bigg]
    \notag \\
    &\quad\mbox{}+\frac{r^{-2}-r^2}{r^{-2}+r^2}.
\end{align}

Let us prepare the initial state of qubit $j$ as
\begin{align}
    \hat{\rho}_{\mathrm{q}j}(0)&=\ket{C_{+}}_{\mathrm{q}j}\bra{C_{+}},
\end{align}
which corresponds to $b_{j,x}(0)=b_{j,y}(0)=0$ and $b_{j,z}(0)=1$.
This leads to $b_{j,x}(t)=b_{j,y}(t)=0$ and
\begin{align}
    b_{j,z}(t)&=\exp{[-\kappa\alpha^2(r^{-2}+r^2)t]}\bigg[1-\frac{r^{-2}-r^2}{r^{-2}+r^2}\bigg]
    \notag \\
    &\quad\mbox{}+\frac{r^{-2}-r^2}{r^{-2}+r^2}.
\end{align}
Thus, the density matrix of qubit $j$ at time $t$ is written as
\begin{align}
    \hat{\rho}_{\mathrm{q}j}(t)&=\frac{1+b_{j,z}(t)}{2}\ket{C_{+}}_{\mathrm{q}j}\bra{C_{+}}
    \notag \\
    &\quad\mbox{}+\frac{1-b_{j,z}(t)}{2}\ket{C_{-}}_{\mathrm{q}j}\bra{C_{-}}
    \notag \\
    &=\frac{\mathcal{N}_+^2+\mathcal{N}_-^2+(\mathcal{N}_+^2-\mathcal{N}_-^2)b_{j,z}(t)}{2}
    \notag \\
    &\quad\times(\ket{\alpha}_{\mathrm{KPO}j}\bra{\alpha}+\ket{-\alpha}_{\mathrm{KPO}j}\bra{-\alpha})
    \notag \\
    &\quad\mbox{}+\frac{\mathcal{N}_+^2-\mathcal{N}_-^2+(\mathcal{N}_+^2+\mathcal{N}_-^2)b_{j,z}(t)}{2}
    \notag \\
    &\quad\times(\ket{\alpha}_{\mathrm{KPO}j}\bra{-\alpha}+\ket{-\alpha}_{\mathrm{KPO}j}\bra{\alpha}),
\end{align}
where in the last equality, we have changed the basis states of qubit $j$ from Eq.~\eqref{eq:basis2} to Eq.~\eqref{eq:basis1}.
When $\alpha$ is large enough that $\mathrm{e}^{-2\alpha^2}\approx0$, $\mathcal{N}_{\pm}\approx1/\sqrt{2}$, and $r\approx1$ are good approximations, we have
\begin{align}
    &\hat{\rho}_{\mathrm{q}j}(t)
    \notag \\
    &\approx\frac{\ket{\alpha}\bra{\alpha}
    +\ket{-\alpha}\bra{-\alpha}
    +\mathrm{e}^{-\gamma t}(\ket{\alpha}\bra{-\alpha}
    +\ket{-\alpha}\bra{\alpha})}{2(1+\mathrm{e}^{-2\alpha^2-\gamma t})},
\end{align}
where $\gamma:=2\kappa\alpha^2$ is the dephasing rate of Kerr-cat qubit $j$ and we have taken normalization into account. 
We thus arrive at Eq.~\eqref{eq:dephased_state1}.

\section{Detuning schedule of the couplers for the \texorpdfstring{$R_{zz}$}{Rzz} gate}
\label{app:detuning_schedule}
\begin{table*}
    \caption{$\alpha_c^{\mathrm{max}}$ that minimizes the infidelity $1-F(\hat{\rho}(t_f),\ket{\Psi^{\mathrm{ideal}}_{\Theta}})$ for each gate time $t_f$.}
    \label{table:parameter2_alpha_c_max}
    \begin{center}
        \begin{tabular}{ccccccccccccccccc}
            \hline\hline
            $t_f$ (ns) & $8$ & $10$ & $12$ & $14$ & $15$ & $16$ & $17$ & $18$ & $19$ & $20$ & $21$ & $22$ & $24$ & $26$ & $28$ & $30$ \\ \hline
            $\alpha_c^{\mathrm{max}}$ & $0.404$ & $0.460$ & $0.496$ & $0.524$ & $0.528$ & $0.524$ & $0.510$ & $0.486$ & $0.468$ & $0.440$ & $0.412$ & $0.408$ & $0.390$ & $0.370$ & $0.336$ & $0.314$ \\ \hline
            $t_f$ (ns) & $32$ & $34$ & $36$ & $38$ & $40$ & $42$ & $44$ & $46$ & $48$ & $50$ & $52$ & $54$ & $56$ & $58$ & $60$ & \\ \hline
            $\alpha_c^{\mathrm{max}}$ & $0.318$ & $0.296$ & $0.280$ & $0.264$ & $0.252$ & $0.244$ & $0.232$ & $0.226$ & $0.230$ & $0.212$ & $0.220$ & $0.202$ & $0.198$ & $0.192$ & $0.188$ & \\
            \hline\hline
        \end{tabular}
    \end{center}
\end{table*}
\begin{figure}
    \centering
    \includegraphics[width=0.3\textwidth]{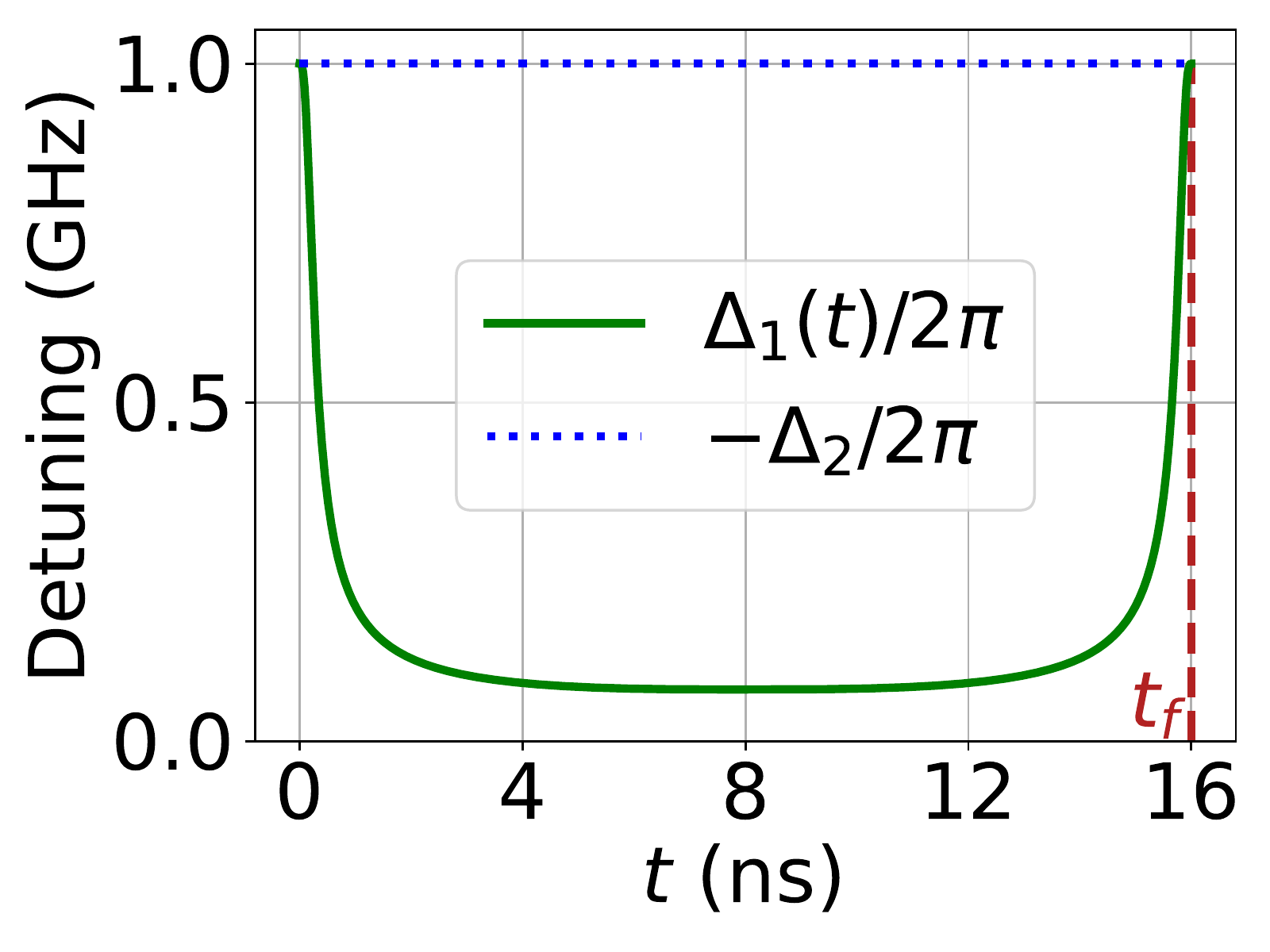}
    \caption{Time dependence of $\Delta_{1}(t)$ and $-\Delta_{2}$ in Eq.~\eqref{eq:detuning1}  for  $t_f=16$\,ns and $\alpha_c^{\mathrm{max}}=0.524$. The other parameters are shown in Table \ref{table:parameter1}.}
    \label{fig:Delta_c12t}
\end{figure}
\TAred{When we evaluate the performance of the $R_{zz}$ gate in Sec.~\ref{subsec:Rzz_gate},} we use a particular detuning schedule of the couplers:
\begin{gather}
    \Delta_{1}(t)
    =\frac{2g\alpha}{\lambda[u(t)]}
    \quad(0\leq t\leq t_f),    \quad
    \Delta_{2}
    =-\frac{2g\alpha}
    {\alpha_c^{\mathrm{min}}},
       \label{eq:detuning1}
    \\
    \lambda(x)
    =\left\{
        \begin{aligned}
            &\alpha_c^{\mathrm{min}}
            +\left(
                \alpha_c^{\mathrm{max}}-\alpha_c^{\mathrm{min}}
            \right)
            \left(
                \frac{x}{T}
                -\frac{1}{2\pi}
                \sin{\frac{2\pi}{T}x}
            \right)
            \\
            &\mbox{}\hphantom{\times\left(
                \frac{x}{T}
                -\frac{1}{2\pi}
                \sin{\frac{2\pi}{T}x}
            \right)}\quad(0\leq x\leq T)
            \\
            &2\alpha_c^{\mathrm{max}}-\alpha_c^{\mathrm{min}}-\left(
                \alpha_c^{\mathrm{max}}-\alpha_c^{\mathrm{min}}
            \right)
            \\
            &\times\left(
                \frac{x}{T}
                -\frac{1}{2\pi}
                \sin{\frac{2\pi}{T}x}
            \right)
            \quad
            (T\leq x\leq 2T)
        \end{aligned}
    \right.,
\end{gather}
where $T=t_f/(\alpha_c^{\mathrm{max}}+\alpha_c^{\mathrm{min}})$ and $u(t)$ is determined from
\begin{align}
    \left\{
        \begin{aligned}
        &\alpha_c^{\mathrm{min}}u(t)
        +\left(
            \alpha_c^{\mathrm{max}}-\alpha_c^{\mathrm{min}}
        \right)
        \left\{
        \frac{u(t)^2}{2T}
        -\frac{T}{(2\pi)^2}
        \right.
        \\
        &\times\left.\left[1-
        \cos{\left(\frac{2\pi}{T}u(t)\right)}
        \right]
        \right\}
        =t\quad(0\leq t\leq t_f/2)
        \\
        &(2\alpha_c^{\mathrm{max}}-\alpha_c^{\mathrm{min}})
        u(t)
        -\left(
        \alpha_c^{\mathrm{max}}-\alpha_c^{\mathrm{min}}
        \right)
        \left\{
        \frac{u(t)^2}{2T}
        +T
        \right.
        \\
        &\mbox{}-\left.\frac{T}{(2\pi)^2}
        \left[1-
        \cos{\left(\frac{2\pi}{T}u(t)\right)}
        \right]
        \right\}
        =t\quad(t_f/2\leq t\leq t_f)
        \end{aligned}.
    \right.
\end{align}
This corresponds to
\begin{align}
    \alpha_{1}(t)=\lambda[u(t)],\quad
    \alpha_{2}=-\alpha^{\rm min}_{\rm c}.
\end{align}
Note that we have
\begin{gather}
    \lambda(2T-x)=\lambda(x)\quad(0\leq x\leq T), \\    u(0)=0,\quad u(t_f/2)=T,\quad u(t_f)=2T,\\
    \lambda[u(0)]=\lambda[u(t_f)]=\alpha^{\rm min}_{\rm c},
    \quad\lambda[u(t_f/2)]=\alpha^{\rm max}_{\rm c},\\
    \Delta_{1}(0)=\Delta_{1}(t_f)= -\Delta_{2}=2g\alpha/\alpha^{\rm min}_{\rm c}=2\pi\times1\,\mathrm{GHz}.
    \label{eq:detuning2}
\end{gather}
The complicated form of $\lambda[u(t)]$ is the result of trial and error.
Predetermined parameters are listed in Table~\ref{table:parameter1}.
In calculations of the infidelity for each gate time $t_f$, we optimize $\alpha^{\rm max}_{\rm c}$ so that we have the minimal infidelity with $\Theta=-\pi/2$.
The used values of $\alpha_c^{\mathrm{max}}$ are listed in Table \ref{table:parameter2_alpha_c_max}.
Typical time dependence of $\Delta_{1}(t)$ is presented in Fig.~\ref{fig:Delta_c12t}.
The point is that we decrease $\Delta_{1}(t)$ as rapidly as possible while avoiding nonadiabatic transitions after $t=0$ and keep it at small value\TAred{s} most of the gate time while keeping $\alpha_c^{\mathrm{max}}$ not too large.
This is because we must satisfy the condition \eqref{eq:condition1}.

\bibliography{aoki3}
\end{document}